\documentclass[11pt,a4paper]{article}
\usepackage[T1]{fontenc}
\usepackage{lmodern}
\usepackage[margin=27mm]{geometry}
\usepackage{amsmath,amssymb,amsthm}
\usepackage{microtype,booktabs,array,graphicx}
\usepackage{placeins,longtable}
\usepackage[numbers,sort&compress]{natbib}
\usepackage{xcolor,tikz}
\usepackage[colorlinks=true,linkcolor=blue!45!black,citecolor=blue!45!black,urlcolor=blue!45!black]{hyperref}
\usetikzlibrary{shapes.geometric}
\newtheorem{theorem}{Theorem}[section]
\newtheorem{proposition}[theorem]{Proposition}
\newtheorem{lemma}[theorem]{Lemma}
\newtheorem{corollary}[theorem]{Corollary}
\theoremstyle{definition}
\newtheorem{definition}[theorem]{Definition}
\theoremstyle{remark}
\newtheorem{remark}[theorem]{Remark}

\DeclareMathOperator{\Res}{Res}
\newcommand{\R}{\mathbb R}
\newcommand{\C}{\mathbb C}

\newcommand{\norm}[1]{\left\lVert#1\right\rVert}
\newcommand{\inner}[2]{\left\langle#1,#2\right\rangle}

\newcommand{\Bobs}{\mathcal B_E}
\title{One Loss Resolves Signed Rank-One Gradient Fibers:\protect\\Exact Decoding on Cycle and Theta Blocks}
\author{Junwen Lou\thanks{Corresponding author. Email: \href{mailto:312219020216@home.hpu.edu.cn}{\texttt{312219020216@home.hpu.edu.cn}}.}, Shuo Tian, Penghui Liu\\[0.4em]
\normalsize Henan Polytechnic University, Jiaozuo, Henan, China}
\date{}

\begin{document}
\maketitle
\begin{abstract}
For masked rank-one least squares with unrestricted signs, gradients
reveal weighted margins and the loss reveals the target's squared norm.
We prove that this one scalar resolves the gradient ambiguity on
bipartite cycle and length-three parallel-path blocks: every target
fixed before sampling and nonzero on cycle edges is identified almost
surely by one legal positive query drawn from an absolutely continuous
law. On six-cycle flowers, open target and query sets retain $3^r$
bounded gradient candidates. A common algebraic witness excludes cross-block energy
cancellation. Cactus masks admit explicit exception certificates and
target-uniform finite-grid bounds; theta blocks admit an existence
guarantee. For nondegenerate cactus reductions, an exact decoder returns
complete actual real lists, with bit cost polynomial in input length,
the cycle-degree product $D$ and the optional budget parameter
$2^\kappa$. Energy-coordinate propagation certifies singletons without
a product join. The product $D$ can grow exponentially in cycle count.
Exponential certificate-degree and output bounds, and
weak NP-hardness at a fixed uniform query, characterize distinct
costs of exact decoding.
\end{abstract}

\section{Introduction}

Objective values and derivatives are naturally evaluated together in
least-squares optimization. What information does the objective value
add to the gradients? For masked rank-one least squares, the gradients
give weighted row and column sums of the unknown observed array, while
the loss gives its total squared norm. With unrestricted signs, those
margins can leave exponentially many rank-one candidates. We show that
one total norm separates them for every target fixed before sampling
and nonzero on cycle edges, almost surely under a random positive query,
on masks built from even
cycles and length-three parallel-path blocks.

The distinction is visible within a single observation model. On
connected six-cycle flowers, open sets of targets and queries have
gradient fibers with exactly $3^r$ actual candidates, all with strictly
bounded factors.
For each fixed target in that family, adding the loss makes the reply
unique almost surely. Thus a scalar already present in the least-squares
evaluation changes the information available for exact reconstruction.

The challenge is to prove separation by this particular scalar.
The weighted-margin kernel has dimension equal to the cycle rank,
also the codimension of the nonzero rank-one torus in observed-entry
space. This suggests a finite candidate fiber. An independent generic
measurement can separate finite fibers \citep{breiding2023}, but here
the norm coordinate is fixed and all gradient rows are linked through
the same query. The result must also hold for every fixed target,
including nongeneric ones. We establish nonvanishing within this
query-linked observation family by ruling out cancellation of the
energy changes across blocks.

\paragraph{Separation and explicit certificates.}
Theorem~\ref{thm:theta-blocks} proves fixed-target identification for
even-cycle and arbitrary-size theta blocks joined at cut vertices.
A scaled inverse-factor specialization makes every false complex
branch have strictly negative real energy change, providing a common
noncancellation witness. On cacti, a private vertex of a leaf cycle
instead drives false branches away from any prescribed compensating
energy. This gives an explicit exception polynomial and a finite-grid
bound uniform over admissible fixed targets
(Theorem~\ref{thm:main}). The witness may depend on the target; the
observer samples independently of it.

\paragraph{Exact reconstruction and its costs.}
The decoder uses only the public mask, query and reply. It isolates
cycle roots, matches total energies and checks finite factorization,
returning the complete actual real list whenever the cycle reduction
is nondegenerate. Rational cactus inputs admit a deterministic bit
bound polynomial in the cycle-degree product $D$, input length and
the optional budget parameter $2^\kappa$
(Theorem~\ref{thm:bitcomplexity}). Energy-coordinate propagation gives
a polynomial-bit sufficient singleton certificate before a product join
(Proposition~\ref{prop:squareclass}); finite theta systems admit
classical rational univariate representations. Certificate-degree,
explicit-output and subset-sum constructions identify three distinct
sources of worst-case cost. Exact computational comparisons test list
coverage and these decoding mechanisms.

\paragraph{Relation to reconstruction and symbolic computation.}
Gradient inversion studies the information retained in released
derivatives \citep{zhu2019,geiping2020}, including provable recovery from
a random neural-network gradient \citep{wang2023}. Here the experimenter
selects a positive factor query for an exactly rank-one observed array.
The resulting margin fibers connect matrix scaling \citep{rothblum1989},
graph toric ideals \citep{ohsugi1999} and quasi-independence likelihood
equations \citep{coons2020,barnhill2024,amendola2021toric}. Their classical
binomials supply the candidate equations; our separation arguments show
how the native norm coordinate distinguishes signed candidates for each
fixed target. Completion and rigidity
\citep{singer2010,kalai2016,kiraly2015,cruickshank2026} address complementary
structured reconstruction questions. The exact decoder combines real
root isolation and algebraic predicates \citep{basu2006,mignotte1992},
rational univariate representations \citep{rouillier1999,demin2025rur}
and sum matching \citep{horowitz1974}. General diagonal-scaling hardness
\citep{khachiyan1996} provides context for the native-interface reduction.

Section~\ref{sec:model} specifies the observation and its quantifiers;
Section~\ref{sec:utility} exhibits the information separation.
Section~\ref{sec:constructive} develops certified decoding, and
Section~\ref{sec:proof} proves global separation.
Sections~\ref{sec:regular} and~\ref{sec:verification} treat regularity
and computational validation, respectively.

\section{Observation model and main results}\label{sec:model}

Let $G=(R\sqcup C,E)$ be a finite simple bipartite graph, with $|R|=m\geq1$
and $|C|=n\geq1$. Isolated vertices and disconnected components are allowed.
For an observed array $B\in\R^E$, its norm and inner product always refer
only to the coordinates in $E$.

\begin{definition}
An \emph{actual rank-one observed target} is an array in
\[
 \Bobs=\{(a_i b_j)_{ij\in E}:a\in\R^m,\ b\in\R^n\}.
\]
The inverse problem recovers this observed array, not its factors or an
unobserved completion. Factors must be finite. The bounded class
$\Bobs^{(1)}$ also requires \mbox{$\norm a,\norm b\leq1$}.
\end{definition}

Actualness matters at zero coordinates: the zero set of cycle equations
can contain points with no finite factorization. Our proof excludes
ambiguities in a larger algebraic set, so such extra points cannot create
a false claim about actual targets.

A legal query has $\norm u,\norm v\leq1$. Its \emph{complete reply}
consists of the scalar loss and both factor-gradient blocks:
\begin{align}
 f_B(u,v)&=\frac12\sum_{ij\in E}(u_i v_j-B_{ij})^2,\label{eq:loss}\\
 (g_u)_i&=\sum_{j:ij\in E}(u_i v_j-B_{ij})v_j, &
 (g_v)_j&=\sum_{i:ij\in E}(u_i v_j-B_{ij})u_i.\label{eq:gradient}
\end{align}
Both gradient blocks and the loss count as \emph{one paired call}.
For a fixed query, set $Q=(u_i v_j)_{ij\in E}$ and $Z_{ij}=u_i v_j B_{ij}$.
Known prediction terms and the gradients determine every row and column
sum of $Z$. In addition,
\begin{equation}
 \norm B^2=2f_B+\norm Q^2-\inner u{g_u}-\inner v{g_v}.
 \label{eq:norm-extraction}
\end{equation}
Indeed each gradient contraction is $\inner Q{Q-B}$.
Thus the target-dependent information is exactly its weighted margins
and \emph{one total norm}. No norm for an individual cycle is supplied.

Let $J_E(q)$ be the Jacobian of the edge-product map
$\Phi_E(u,v)=(u_iv_j)_{ij\in E}$ at $q=(u,v)$. For an observed array $C$,
after removing known prediction terms, the complete observation is
$(J_E(q)^T C,\norm C^2)$. Thus it is a linear projection of the lifted
target $(C,\norm C^2)$ whose gradient rows are linked through $q$.

A block is a maximal biconnected subgraph; bridges are single-edge blocks.
A block is nontrivial when it contains a cycle. Let $\Theta_p$ denote
$p\geq3$ internally disjoint paths of length three between a row vertex
and a column vertex. Its six-cycles share whole paths. Different blocks
may have different path counts and may meet at arbitrary cut vertices.

\begin{theorem}[Cycle and theta blocks]\label{thm:theta-blocks}
Let every nontrivial block of a bipartite mask be an even cycle or
$\Theta_p$ for some $p\geq3$, without a bound on the number of paths.
For every actual real target fixed before
sampling and nonzero on cycle edges, one legal positive complete query
identifies it almost surely under any absolutely continuous law supported
on legal positive queries.
Its exceptional queries lie in the zero set of a nonzero polynomial
depending on the target. Competitors may contain zeros, and bridge
entries of the fixed target may be zero.
\end{theorem}

Section~\ref{sec:theta} proves this result by a common algebraic
specialization. On cacti the guarantee has an explicit grid certificate.

A graph is a \emph{cactus} if every edge belongs to at most one simple
cycle. Suppose now that $G$ is a cactus. Write its cycles as $C_\ell$, of length
$2k_\ell$, for $\ell=1,\ldots,r$, and define
\begin{align}
 d_\ell&=\begin{cases}k_\ell,&k_\ell\text{ odd},\\k_\ell-1,&k_\ell\text{ even},\end{cases}
 &D&=\prod_{\ell=1}^r d_\ell,& r_+&=\#\{\ell:d_\ell>1\},\label{eq:degrees}\\
 L_G&=2\sum_{\ell=1}^r d_\ell+4r_+D, &
 \widetilde L_G&=2\sum_{k_\ell\text{ even}}d_\ell+4r_+D-2r_+.
 \label{eq:bound}
\end{align}
Empty products equal one and empty sums equal zero.

\paragraph{Notation.}
The mask is $G$ and the observed target is $B$; $N$ denotes grid size,
whereas $\mathcal N$ denotes extracted squared norm. Cycle degrees are
$d_\ell$, with product $D$; reduced degrees are $e_\ell=d_\ell-1$.
In the arithmetic appendices, $Q_{\rm den}$ clears public denominators,
$Q_{\rm aff}$ clears affine-edge denominators and $Q_B$ clears target
entries. Hardness offsets $c_\ell^-,c_\ell^+$ and subset target
$T_{\rm ss}$ are separate from these degrees and denominator bounds.

\begin{theorem}[Cactus grid certificate]\label{thm:main}
Fix $B\in\Bobs$ before sampling the query, and assume every entry of $B$
on a cycle is nonzero. Independently sample all $x_i,y_j$ uniformly from
$\{1,\ldots,N\}$, and query
\begin{equation}
 u_i=\frac1{m x_i},\qquad v_j=\frac1{n y_j}.\label{eq:grid-query}
\end{equation}
Except with probability at most $\min\{1,\widetilde L_G/N\}$, the complete reply
uniquely determines $B$ among all targets in $\Bobs$, including competitors
with zero entries. Forest-edge entries of $B$ may be zero.
For an absolutely continuous distribution supported on legal positive
queries, uniqueness holds almost surely for each fixed such $B$.
\end{theorem}

The failure coefficient depends only on the mask, so the same bound
holds separately for every admissible fixed target. The exceptional
polynomial itself depends on $B$. Forests have $\widetilde L_G=0$ and
are already recovered by gradients. Target factors may be unbounded,
whereas the queries satisfy $\norm u\leq1/\sqrt m$ and similarly for $v$.
Section~\ref{sec:proof} constructs this cactus certificate by private-vertex
escape and explicit degree counting.

\begin{remark}[Fixed targets and simultaneous recovery]\label{rem:quantifier}
On the full $2\times2$ mask, every legal positive query admits two
strictly bounded nonzero targets with the same complete reply. Set
\[
 a=\tfrac12(u_2,-u_1),\qquad b=\tfrac12(v_2,-v_1),
 \qquad B=ab^T,\quad C=-B.
\]
Both factor norms are at most $1/2$. Orthogonality gives
$Bv=B^Tu=0$ and the same identities for $C$. Their common reply is
\[
 g_u=\norm v^2u,\qquad g_v=\norm u^2v,\qquad
 f=\tfrac{17}{32}\norm u^2\norm v^2.
\]
Here the two targets depend on the query. Theorem~\ref{thm:main}
fixes the target before sampling; its uniform failure bound retains
that order of quantifiers.
\end{remark}

\section{Information separation and its numerical boundary}\label{sec:utility}
The loss distinguishes a fixed target even when its gradient fiber is
exponentially large throughout an open set of targets and queries.
The following result uses the same target class and observation interface
on both sides of this comparison.

\begin{corollary}[Exponential separation on open sets]\label{cor:exponential-separation}
For every $r\geq1$, the connected flower of $r$ six-cycles has a
nonempty relatively open set $\mathcal U_r$ of all-edge-nonzero actual
targets and a nonempty open set $\mathcal V_r$ of legal positive queries
such that, for every $(B,q)\in\mathcal U_r\times\mathcal V_r$, the
entire gradient fiber consists of exactly $3^r$ distinct actual targets.
All these candidates admit factors of norm strictly below one.
For each fixed $B\in\mathcal U_r$, the complete reply identifies $B$
among all actual targets for almost every legal positive query.
\end{corollary}
\begin{proof}
The six-cycle construction of Proposition~\ref{prop:output-size} has
three simple real branches per cycle and strict factor-norm bounds.
The joint implicit-function theorem and continuous factor propagation
preserve all $3^r$ branches on a product neighborhood. Theorem~\ref{thm:main}
then separates each fixed target by the loss. Appendix~\ref{app:matched-information}
gives the complete continuation proof and an explicit two-cycle example.
\end{proof}

Exact identification need not provide a uniform numerical margin.
In the two-cycle example, a distinct bounded gradient-compatible array
stays separated from the target while its loss gap vanishes as the query
approaches a collision. Equation~\eqref{eq:conditioning-gap} gives both
gaps explicitly. This is a branch-separation effect, distinct from local
tangent conditioning.

\section{Exact public-input list decoding}\label{sec:constructive}

The margin equations reduce reconstruction to cycle parameters; the loss
selects tuples by their total energy. The remaining step is to certify
that each selected observed array has finite rank-one factors. We first
give this support test, then combine it with exact root isolation,
energy matching and arithmetic singleton certificates.

\subsection{Finite factorization and the optional factor budget}

Algebraic cycle equations alone do not enforce finite factorization. For a proposed real array $C$, call a vertex active if it
is incident to a nonzero entry, and let $H$ be its nonzero-edge subgraph.
Propagate factors along a spanning tree of each component of $H$, starting
one factor at one. Check the remaining nonzero edges against those factors.

\begin{lemma}[Support and factor budget]\label{lem:actual}
An array $C$ is actual if and only if this propagation is consistent and
no zero edge has two active endpoints. If these conditions hold, let
$a^{(j)},b^{(j)}$ be the propagated factors of the nontrivial components
of $H$, and put $A_j=\norm{a^{(j)}}^2$, $B_j=\norm{b^{(j)}}^2$. Then
\begin{equation}
 C\in\Bobs^{(1)}\quad\Longleftrightarrow\quad
 \sum_j\sqrt{A_jB_j}\leq1.\label{eq:factor-budget}
\end{equation}
The empty sum is zero.
\end{lemma}
\begin{proof}
An active vertex must have a nonzero factor, so a zero edge between two
active vertices is impossible. Consistency is necessary on each component.
Conversely, use the propagated factors on active vertices and set inactive
factors to zero. All observed edges then have their prescribed values.
Any other nonzero factorization of component $j$ has factors
$c_j a^{(j)},b^{(j)}/c_j$ for a nonzero scalar $c_j$; this follows by
propagating ratios on its connected graph. Consequently Cauchy--Schwarz gives
\[
 \Bigl(\sum_j A_j c_j^2\Bigr)
 \Bigl(\sum_j B_j/c_j^2\Bigr)
 \geq\Bigl(\sum_j\sqrt{A_jB_j}\Bigr)^2.
\]
If both factor norms are at most one, the right-hand sum is at most one.
For sufficiency choose $c_j^2=\sqrt{B_j/A_j}$, making both global squared
factor norms equal to that sum. Isolated factors are set to zero.
\end{proof}

\begin{proposition}[Public-input enumeration]\label{prop:decoder}
Given a consistent complete reply at a positive query off the
leading-coefficient exceptions of Theorem~\ref{thm:main}, the following
procedure enumerates every actual gradient-consistent target before its
final norm test, using at most $D$ cycle tuples. The unit-factor
restriction can optionally be imposed by \eqref{eq:factor-budget}.
\end{proposition}
\begin{proof}
Apply Algorithm~1 below before its norm test. Off the leading exceptions,
the cycle polynomials have degrees $d_\ell$, giving at most $D$ tuples.
Lemmas~\ref{lem:incidence} and~\ref{lem:actual} give completeness and the
exact actualness and optional budget tests.
\end{proof}

Exact root isolation and algebraic comparisons make the construction
effective for rational or algebraic public data. Numerical variants
must retain an unresolved status for undecided degrees or equalities.
An empty mask has one empty observed array.

\subsection{Certified list decoding and its cost}\label{sec:algorithm}

The decoder certifies the supplied reply directly. It checks the
actual cycle equations, so the observer can establish uniqueness without
constructing the target-dependent exception polynomial. Identically zero
cycle equations receive a separate return status.

\begin{samepage}
\textbf{Algorithm 1: Exact cactus list decoder.}
Input: a cactus mask, a positive rational query, and a rational complete
reply; optionally impose the unit-factor bound.
\begin{enumerate}
\item Extract the weighted margins and total norm by
\eqref{eq:norm-extraction}. Solve the incidence system over $\mathbb Q$.
If it is inconsistent, return an empty list with that certificate.
\item Find the edge-disjoint cycles and alternating kernel vectors.
Choose a rational particular solution $Z_0$, for example by setting
nonforest edges to zero and removing forest leaves. For each cycle construct
its binomial $p_\ell(t_\ell)$ on the affine margin fiber.
\item A nonzero constant polynomial gives an empty list. Otherwise, if
some $p_\ell$ is identically zero, return \textsc{Degenerate}, without a
uniqueness claim. For every nonconstant polynomial, take
its square-free part and isolate all distinct real roots exactly.
\item Enumerate the Cartesian product of those root sets. Undo the
query weights and first reject candidates failing total-norm equality.
For the survivors test support propagation and, if requested,
\eqref{eq:factor-budget}. Retain exactly the candidates passing all tests.
\item Return the list together with the univariate root representations
and exact tests. Label a singleton \textsc{Unique}; label a longer list
\textsc{Ambiguous}. For a forest the Cartesian product has one empty tuple;
for an empty mask return the unique empty observed array when the reply
is consistent.
\end{enumerate}
\end{samepage}

\begin{theorem}[Complete decoding and checked uniqueness]\label{prop:certificate}
Whenever Algorithm~1 returns a list rather than \textnormal{\textsc{Degenerate}},
that list contains exactly all actual targets giving the supplied reply,
with the optional factor bound if imposed. Thus a singleton certifies
uniqueness for that particular reply, including queries not covered by
the sufficient exceptional-set test. A degree drop alone does not invalidate
the algorithm: the actual nonzero polynomial degree is used.
\end{theorem}
\begin{proof}
The incidence kernel describes every gradient-consistent observed array.
Every actual target satisfies the cycle binomials. Root isolation exhausts
their real zeros, with multiplicities removed but no distinct roots lost.
Lemma~\ref{lem:actual} rejects exactly the nonactual or out-of-budget
candidates. The norm test is equivalent to the remaining scalar loss
constraint. Distinct tuples give distinct arrays because the alternating
kernel vectors are linearly independent. Identically zero polynomials
are excluded before any finite-list claim is made.
\end{proof}

The \textsc{Degenerate} status has a geometric reason: a reply can have
an infinite actual fiber. On a flower of
$r\geq2$ four-cycles, a uniform query and zero weighted margins leave
one alternating edge parameter $c_\ell$ free on each cycle. Every tuple
with all $c_\ell\ne0$ is actual: take the shared row factor to be one,
each private row factor to be minus one, and the two column factors
to be $c_\ell,-c_\ell$. Fixing the loss imposes
$4\sum_\ell c_\ell^2=\mathcal N$, which has an $(r-1)$-dimensional
actual part when $\mathcal N>0$. A finite-list output cannot represent
such a reply.

Complete root lists and exact predicates certify the output, either
over all actual targets or with the optional unit-factor restriction.
For rational targets satisfying Theorem~\ref{thm:main}, independent queries
with $N>\widetilde L_G$ may be repeated until a singleton appears.
Their expected number is at most $(1-\widetilde L_G/N)^{-1}$; every returned
answer is correct. This Las Vegas bound counts calls.
Sections~\ref{sec:complexity} and~\ref{sec:bitcomplexity} and
Appendices~\ref{app:bitcomplexity} and~\ref{app:certificate-heights}
give arithmetic, coefficient-height and bit-cost bounds.

\subsection{Combining exact cycle energies}
Total energy is additive over edge-disjoint cycles. After root isolation,
classical meet-in-the-middle matching \citep{horowitz1974} combines two
partial energy lists and retains every equal-value pair, before support
and optional budget filtering. Proposition~\ref{prop:mitm} gives exact operation and storage counts.

\subsection{Rational heights and an exact energy join}\label{sec:complexity}

We first bound the heights of the decoder's public-input cycle
equations. This supplies the rational encoding needed for exact
root isolation and energy comparison.

\begin{proposition}[Cycle coefficient height]\label{prop:height}
Put $V=m+n\geq1$, $e=|E|$, and $M_{\rm pub}=2V+1$. Suppose every rational
coordinate of $u,v,g_u,g_v,f$ has numerator of absolute value less than
$2^b$ and positive denominator less than $2^b$, with $b\geq1$.
Choose the particular weighted array by setting nonforest edges to zero
and solving the forest margins. Define
\[
 H_{\rm cyc}=4(M_{\rm pub}+1)b+\lceil\log_2(\max\{1,V\}(e+1))\rceil+2.
\]
For a cycle of length $2k$, clearing denominators and taking the primitive
integer part gives a polynomial whose coefficient bit lengths are at most
$k(H_{\rm cyc}+1)+2$. Forest margin solving takes $O(V+e)$ rational operations
after a spanning forest has been chosen.
\end{proposition}
\begin{proof}
Let $Q_{\rm den}$ be the product of the public coordinate denominators, so
$Q_{\rm den}<2^{M_{\rm pub}b}$. A weighted margin is
$u_i^2\sum_{j:(i,j)\in E}v_j^2-u_i(g_u)_i$, or its column analogue.
It has denominator dividing $Q_{\rm den}^4$ and absolute value at most
$(e+1)2^{4b}$. Remove forest leaves successively: a leaf edge takes
the remaining margin at that vertex, which is then subtracted at its
neighbor. Every resulting edge is a signed sum of at most $V$ original
margins. Thus we may write $Z_{0,e}=a_e/Q_{\rm den}^4$ with
$|a_e|<2^{H_{\rm cyc}}$ and $Q_{\rm den}^4<2^{H_{\rm cyc}}$. Clearing the denominator $Q_{\rm den}^{4k}$ in a cycle
binomial gives the difference of two products of $k$ linear factors
$a_e\pm Q_{\rm den}^4t$. Every coefficient is bounded by
$2\binom{k}{j}2^{kH_{\rm cyc}}\leq2^{k(H_{\rm cyc}+1)+1}$.
Dividing by integer content cannot increase this height. Each forest
edge is removed once, giving the stated operation count.
\end{proof}

Total energy separates over the edge-disjoint cycles even though
actualness and the optional factor budget need not separate. Write
\[
 \mathcal N(t)=\mathcal N_{\rm forest}
                  +\sum_{\ell=1}^r h_\ell(t_\ell),\qquad
 h_\ell(t)=\sum_{e=(i,j)\in C_\ell}
           \frac{(Z_{0,e}+n_{\ell,e}t)^2}{u_i^2v_j^2}.
\]
Each $h_\ell$ is quadratic with rational coefficients. After isolating
the cycle roots, one can therefore filter by energy \emph{before}
support propagation. This uses classical meet-in-the-middle sum
matching \citep{horowitz1974}, applied here with exact real algebraic
energies rather than integer weights.

\begin{proposition}[Exact energy matching]\label{prop:mitm}
Let $n_{\ell,\mathbb R}$ be the number of distinct real roots of the
nonzero cycle polynomial $p_\ell$, and put
$D_{\mathbb R}=\prod_\ell n_{\ell,\mathbb R}\leq D$.
Partition the cycles into $\mathcal I_L,\mathcal I_R$ and put
$D_L=\prod_{\ell\in\mathcal I_L}n_{\ell,\mathbb R}$,
$D_R=\prod_{\ell\in\mathcal I_R}n_{\ell,\mathbb R}$, so $D_LD_R=D_{\mathbb R}$.
Let $K_{\rm match}$ be the number of complete
root tuples passing the energy equality, before actualness or budget
filtering. After root isolation the complete actual list can be found in
\[
 O\bigl(r(D_L+D_R)+D_L\log(2+D_L)+D_R\log(2+D_R)+K_{\rm match}(V+e)\bigr)
\]
exact algebraic arithmetic, square-root and order/equality operations.
Partial-list storage uses $O((r+1)(D_L+D_R))$ algebraic numbers and
indices, excluding output.
\end{proposition}
\begin{proof}
Enumerate the partial sums on $\mathcal I_L$ and the required complementary sums
$\mathcal N-\mathcal N_{\rm forest}-\sum_{\ell\in\mathcal I_R}h_\ell$ on $\mathcal I_R$.
Sort both lists by exact real order and merge them. For equal values,
output the Cartesian product of the two entire equal-value blocks;
retaining only one representative would lose solutions. Every matching
pair determines one full tuple, and the partition makes this
correspondence bijective. Enumeration, sorting and the monotone merge
give the counts, with $O(r)$ coordinates per output tuple. Applying
Lemma~\ref{lem:actual} to exactly these tuples preserves the complete list;
its $O(V+e)$ cost also absorbs reconstruction of each tuple.
\end{proof}

When the cycle root counts admit a balanced partition, both partial
lists have size close to $\sqrt{D_{\mathbb R}}$, the square root of the
real-root tuple count. The displayed bound records the actual partition
sizes and all $K_{\rm match}$ matches, including repeated equal-energy
blocks. A cycle with a dominant root count can prevent balance.
These are exact-operation counts; the bit bound in
Theorem~\ref{thm:bitcomplexity} also accounts for the algebraic fields
needed to compare partial sums.

\subsection{A polynomial-bit energy-coordinate certificate}\label{sec:squareclass}

Quadratic cycle energies admit rational coordinates that can certify a
singleton before a product join. Two positive nonsquare rationals have
the same square class if their ratio is a rational square. Integer
square roots test this equivalence without integer factorization. We
use the classical rational-independence framework for radicals
\citep[Section 4.1]{bloemer1993} and apply finite-domain bounds propagation \citep{choi2004bounds}
to these exact coordinates.

\begin{proposition}[Energy-coordinate propagation]\label{prop:squareclass}
Suppose the public input is rational, every cycle polynomial is nonzero
of degree at most three, and all its irreducible factors over $\mathbb Q$
have degree at most two. Write each distinct real root's energy as
\[
 e_{\ell j}=a_{\ell j}+b_{\ell j}\sqrt{d_{\ell j}},
 \qquad a_{\ell j},b_{\ell j},d_{\ell j}\in\mathbb Q,
\]
with $d_{\ell j}>0$ nonsquare when $b_{\ell j}\ne0$.
Choose representatives $\delta_1,\ldots,\delta_s$ for all nonrational
energy classes. Let $v_{\ell j}\in\mathbb Q^{s+1}$ represent $e_{\ell j}$
in the coordinates $1,\sqrt{\delta_1},\ldots,\sqrt{\delta_s}$, and put
$\tau=(\mathcal N-\mathcal N_{\rm forest},0,\ldots,0)$.
Initially let $J_\ell$ contain all real roots on cycle $\ell$. Repeatedly
delete $j\in J_\ell$ whenever, for some coordinate $k$,
\[
 \tau_k-v_{\ell j,k}\notin
 \left[\sum_{a\ne\ell}\min_{b\in J_a}v_{ab,k},\quad
       \sum_{a\ne\ell}\max_{b\in J_a}v_{ab,k}\right].
\]
Stop at an empty domain or a pass with no deletion. No complete-reply
candidate is deleted. The procedure and its deletion certificate have
polynomial bit cost and size in the rational input length and graph size.

An empty domain proves the list empty. If every domain has one root and
these roots are rational, a rational reply and support test decides
whether the complete actual list is empty or a singleton, in polynomial
bit time. The candidate class is all actual real targets, without a
factor-norm restriction.
\end{proposition}
\begin{proof}
The required norm and bridge energy are rational. Distinct positive
rational square classes, together with $1$, have linearly independent
square roots over $\mathbb Q$. In their finite multiquadratic extension,
these classes give distinct characters of the sign-change Galois group.
Multiplying a rational relation by a character and averaging isolates its
coefficient; multiplicative relations such as $2,3,6$ are permitted.
Thus energy equality is exactly $\sum_\ell v_{\ell j_\ell}=\tau$.

For a surviving complete tuple, its other-cycle contribution to each
coordinate lies between the displayed sums of minima and maxima.
A root failing that necessary condition belongs to no such tuple.
Induction over deletion passes therefore preserves every complete-reply
candidate, before actualness filtering. Each deletion record specifies
its cycle, root, coordinate and rational interval; a verifier checks the
inequality directly from the domains preceding that pass.

Rational polynomial factorization has polynomial bit complexity
\citep{lenstra1982}. At degree at most three, rational roots, quadratic
formulas and their quadratic energies have polynomial-size rational
representations. There are at most $3r$ roots and $3r$ classes. Pairwise
rational-square ratio tests construct the coordinates with polynomially
many integer square-root operations. Each successful pass deletes a root,
so at most $3r$ such passes occur. With totals of the coordinate extrema,
one pass uses $O(r^2)$ rational operations. Sums of at most $r$ original
coordinates retain polynomial bit length. Propagation therefore takes
$O(r^3)$ rational operations after coordinate construction, with at most
$3r$ deletion records of polynomial size. When $r=0$, use the final test.

Rational singleton roots give one rational array after undoing the
weights. Recomputing the loss and both gradients and applying
Lemma~\ref{lem:actual} uses rational equalities and zero tests of
polynomial bit size. Every complete-reply candidate must be this array,
and it is retained exactly when those tests pass.
\end{proof}

Private-class deletion is a special case: a class occurring on only one
cycle has other-cycle interval $[0,0]$, excluding its nonzero coefficients.
Propagation also uses the rational coordinate and shared classes. For
example, two cycles with radical coefficients $\{0,\pm1\}$ and
$\{0,\pm2\}$ in the same class must both choose zero. Equal opposing
coefficients can instead survive. The supplied reply therefore determines whether this sufficient
certificate applies; surviving multiple domains pass to the general
decoder.

Integer-lattice energies also admit a pseudopolynomial count without
printing all matching tuples; Appendix~\ref{app:lattice} gives the exact
criterion and its proof.
\paragraph{An intrinsic cost of complete lists.}
The energy join reduces combination work when few sums match. Its output
can nevertheless be exponential even on connected masks, with every
candidate strictly inside the unit-factor class.

\begin{proposition}[Exponential complete-reply lists]\label{prop:output-size}
For every $r\geq1$, a connected flower of $r$ six-cycles admits a legal
rational positive query whose gradient-only fiber contains exactly $3^r$
actual targets. For each $j\in\{0,\ldots,r\}$, a rational loss value
selects exactly $\binom rj2^j$ of them. All these targets have nonzero
entries and admit factors with both norms strictly less than one; the
counts are unchanged by imposing the unit-factor constraint. With
$D=3^r$, some complete-reply list has at least $D/(r+1)$ targets.
Writing such a list explicitly as observed arrays requires $\Omega(D)$
scalar output entries.
\end{proposition}
\begin{proof}
Identify row vertex $0$ of $r$ six-cycles, using edge order
$(00,10,11,21,22,02)$ on each cycle. Thus $m=2r+1$, $n=3r$, and
$|E|=6r$. Put $\lambda=1/(256r)$ and
\[
 C(s)=\lambda(1+s,1-s,1+s,1-s,-2+s,-2-s).
\]
Take every query coordinate to be $q=1/(4r)$. Both query blocks have
squared norm at most $3/16$. On each cycle, the row sums are
$\lambda(-1,2,-1)$ and the column sums are $\lambda(2,2,-4)$,
independent of $s$. The shared row contributes $-r\lambda$.
The incidence kernel has exactly one alternating direction per cycle
(Lemma~\ref{lem:incidence}), so every gradient-compatible array is of
this form with one independent parameter per cycle.

The cycle binomial is
\[
 \lambda^3\bigl((1+s)^2(-2+s)-(1-s)^2(-2-s)\bigr)
       =2\lambda^3s(s^2-3).
\]
Hence every actual candidate has $s\in\{0,\sqrt3,-\sqrt3\}$ on each
cycle. Conversely, all $3^r$ choices are actual. To see this with uniform
factor bounds, set the shared raw row factor to one. For $s=0$, use raw
private row factors $(1,1)$ and column factors $(1,1,-2)$. For
$s=\pm\sqrt3$, use
\[
 (a_1,a_2)=(s-2,7-4s),\qquad
 (b_0,b_1,b_2)=(1+s,-5-3s,-2-s).
\]
Their edge products are $C(s)/\lambda$, and all cycles share the same
row factor. Scale both global factor vectors by $\sqrt\lambda$.
Since $|a_1|<4$, $|a_2|<14$ and $|b_0|<3$, $|b_1|<11$,
$|b_2|<4$ for every branch, their squared norms satisfy
\[
 \norm a^2<\frac{1+212r}{256r}\leq\frac{213}{256}<1,
 \qquad \norm b^2<\frac{146}{256}<1.
\]

Each cycle has squared observed norm $\lambda^2(12+6s^2)$.
Prescribe total norm $\lambda^2(12r+18j)$. A candidate has this norm
precisely when $j$ cycles use a nonzero parameter. Choosing the cycles
and their signs gives exactly $\binom rj2^j$ targets. All margins and
prescribed norms are rational, hence so are the gradients and loss at
every layer. Every candidate entry is nonzero.

The layer sizes sum to $3^r=D$, so some layer has at least $D/(r+1)$
targets. Each observed array has $6r$ entries. Its explicit list therefore
writes at least $6rD/(r+1)\geq3D$ scalars, proving the output bound.
\end{proof}

Exceptional replies can therefore require exponential output even
within the connected, bounded, nonzero-cycle target class. This cost
comes from writing the list. The reduction in
Theorem~\ref{thm:singleton-hardness} identifies a second source of cost:
selecting a branch can encode subset sum even when the final list is
a singleton.

\subsection{A deterministic bit implementation}
\label{sec:bitcomplexity}

Rational root intervals and an algebraic-integer separation bound
implement every zero and sign test at finite precision.

\begin{samepage}
\begin{theorem}[Rational-input bit implementation]\label{thm:bitcomplexity}
Let $V=m+n$, $e=|E|$, and let $b\geq1$ bound the numerator and denominator
bit lengths of all rational public query and reply coordinates. Suppose
the cycle polynomials in Algorithm~1 are not identically zero. Use the
forest particular solution, and the degree product $D$ from
\eqref{eq:degrees}. Test total energy before the optional budget.
Let $\kappa$ be the maximum number of nontrivial nonzero-edge support
components among the actual energy-compatible candidates before budget
filtering, or zero if there are none; always $\kappa\leq\min\{m,n,e\}$.
There is a deterministic implementation returning the exact list and
isolating-root representations in
\begin{equation}
 \bigl(D\,2^\kappa(V+e+1)(b+1)\bigr)^{O(1)}
 \quad\text{bit operations}.\label{eq:bitbound}
\end{equation}
Without the factor bound, the factor $2^\kappa$ is omitted. The exponent
in $O(1)$ is absolute. The same qualitative bound holds for the exact
energy-join variant, without asserting a smaller bit cost for that variant.
A nonzero constant cycle equation gives an empty list during preprocessing.
\end{theorem}
\end{samepage}
\begin{proof}
Proposition~\ref{prop:height} bounds all rational preprocessing heights.
Cross-multiplication makes support tests polynomial in the affine edge
forms. For the budget, the positive product of squared nonzero edges
clears all propagation denominators. The remaining sum of $h$ square
roots has algebraic degree at most $D2^h$. A common integer scaling and
the field norm give an explicit nonzero separation bound with precision
polynomial in $D2^h,V+e,b$. Rational interval refinement then decides
zero and sign, including the zero case. Appendix~\ref{app:bitcomplexity}
gives the scaling, precision and bit-height accounting. Comparisons of two
partial energy sums may involve two root tuples, with degree at most
$D^2$; this remains polynomial in $D$ in the same argument.
\end{proof}

The component parameter $\kappa$ is measured after the energy test and
is bounded by $\min\{m,n,e\}$. For fixed cycle count the unbounded
decoder has polynomial bit complexity; the optional bounded decoder has
this property when $\kappa$ is fixed as well. These are implementation
bounds, separate from the timings of the Python reference code.

\subsection{A fixed-query hardness barrier}\label{sec:hardness}

A small output need not make the energy selection easy. The following
result concerns the binary length of a supplied rational public input.

\begin{theorem}[A parsimonious energy encoding]\label{thm:singleton-hardness}
There is a polynomial-time transformation from positive-integer
\textnormal{\textsc{Subset Sum}} to complete-reply feasibility on a connected flower
of six-cycles at the fixed positive query $u_i=1/m$, $v_j=1/n$.
Its actual gradient fiber consists of exactly $2^r$ rational arrays,
all edge-nonzero and with strictly subunit factors. Subsets of the
prescribed sum correspond bijectively to its complete-reply candidates.
Consequently feasibility is weakly NP-hard. If a deterministic algorithm
recovers the sole array in polynomial bit time on every singleton
input of this family, then $\mathrm{NP}=\mathrm{RP}$.
\end{theorem}

The construction uses cycle roots $0,q_\ell,p_\ell$, where $p_\ell-q_\ell=1$. The zero
root fails actualness, and the two actual branches have an arbitrary
prescribed integer energy difference after rational scaling. The singleton consequence uses the classical isolation lemma
\citep{mulmuley1987}; its full proof is in Appendix~\ref{app:hardness}.
The reduction transports the binary-weight obstruction of subset sum
\citep{karp1972}. For unary weights its energy selection has a
pseudopolynomial dynamic program, as in Section~\ref{sec:lattice}.
This establishes a weak hardness barrier; strong NP-hardness of the
general native problem is open. Meet-in-the-middle instead has
$O^*(2^{r/2})$ time and storage on this binary-branch family
\citep{horowitz1974}.

The reduction fixes the uniform query independently of the supplied
reply. It establishes worst-case cost for uniquely feasible inputs;
this query has measure zero under an absolutely continuous law.
Average-case and smoothed complexity under random queries remain open
and require an analysis of root isolation, algebraic comparison and
energy matching under a specified distribution.

\section{Fixed-target global separation}\label{sec:proof}

The proof first constructs an explicit cactus certificate by separating
the energy of a changed leaf cycle from the other cycles. A common
normalization witness then handles theta blocks. We work over the complex
algebraic candidate fiber; complex squared norms denote polynomial sums
of squares without conjugation. Targets and sampled queries remain real.

\subsection{The classical margin fiber}

Let $A$ be the unsigned vertex-edge incidence matrix of $G$. The margin
equations are $AZ'=AZ$. Orient every cycle so that its alternating vector
$n_\ell$ has entries $+1,-1$ in turn and is zero elsewhere.

\begin{lemma}\label{lem:incidence}
The vectors $n_1,\ldots,n_r$ form a basis of $\ker A$. Hence every ambient
gradient candidate has
\begin{equation}
 Z'=Z+\sum_{\ell=1}^r t_\ell n_\ell.\label{eq:fiber}
\end{equation}
Its bridge coordinates agree with $Z$.
\end{lemma}
\begin{proof}
Changing the signs of the rows indexed by one bipartition converts $A$
into an oriented incidence matrix. Its rank is the number of vertices
minus the number of connected components, with isolated vertices included.
Consequently its kernel dimension is the cycle rank, which is $r$ on a
cactus. The displayed cycle vectors lie in the kernel and have disjoint
edge supports, so they are independent. A bridge lies on none of them.
\end{proof}

For a cycle, write $C_\ell^+$ and $C_\ell^-$ for its two alternating
matchings. Every actual rank-one candidate satisfies the classical binomial
\begin{equation}
 F_\ell(t)=\prod_{e\in C_\ell^+}(z_e+t)
              -\prod_{e\in C_\ell^-}(z_e-t)=0.
 \label{eq:cycle-polynomial}
\end{equation}
The true target satisfies $F_\ell(0)=0$. Let $G_\ell=F_\ell/t$.
The degree of $F_\ell$ is at most $d_\ell$. For odd $k_\ell$ its leading
coefficient is two; for even $k_\ell$ the degree-$k_\ell$ terms cancel
and the degree-$(k_\ell-1)$ coefficient is $\sum_{e\in C_\ell}z_e$.
Under the target hypothesis, that sum is a nonzero polynomial in the query:
distinct edges have distinct query monomials and nonzero fixed coefficients.

By disjointness of cycle edge supports, the norm change is
\begin{align}
 \norm {B'}^2-\norm B^2&=\sum_{\ell=1}^r E_\ell(t_\ell),\label{eq:global-energy}\\
 E_\ell(t)&=A_\ell t^2+2T_\ell t,&
 A_\ell&=\sum_{ij\in C_\ell}\frac1{u_i^2v_j^2},&
 T_\ell&=\sum_{ij\in C_\ell}\frac{n_\ell(ij)B_{ij}}{u_i v_j}.
 \label{eq:local-energy}
\end{align}
If a cycle changes, its parameter is nonzero and must be a root of
$G_\ell$. There are at most $D$ complex cycle-consistent tuples off the
leading-coefficient exceptions. This finite count does not itself show
that their total energies differ: terms in \eqref{eq:global-energy} can
cancel.

\subsection{A private vertex excludes finite compensating energies}

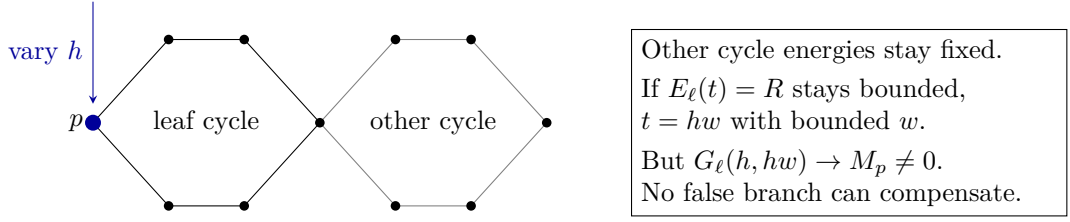
\begin{figure}[tb]
\centering
\begin{tikzpicture}[font=\small,>=stealth]
\coordinate (shared) at (0,0);
\foreach \name/\xx/\yy in {a/-1/1.1,b/-2/1.1,p/-3/0,c/-2/-1.1,d/-1/-1.1,e/1/1.1,f/2/1.1,g/3/0,i/2/-1.1,j/1/-1.1}{\coordinate (\name) at (\xx,\yy);}
\draw (shared)--(a)--(b)--(p)--(c)--(d)--(shared);
\draw[gray] (shared)--(e)--(f)--(g)--(i)--(j)--(shared);
\foreach \name in {shared,a,b,c,d,e,f,g,i,j}{\fill (\name) circle (1.8pt);}
\fill[blue!60!black] (p) circle (3pt);
\node[left] at (p) {$p$};\node at (-1.5,0) {leaf cycle};
\node at (1.5,0) {other cycle};
\draw[->,blue!60!black] (-3,1.6)--node[left]{vary $h$}(-3,.25);
\node[draw,align=left,text width=55mm] at (7,0) {
Other cycle energies stay fixed.\\[3pt]
If $E_\ell(t)=R$ stays bounded,\\
$t=hw$ with bounded $w$.\\[3pt]
But $G_\ell(h,hw)\to M_p\neq0$.\\
No false branch can compensate.};
\end{tikzpicture}
\caption{Private-vertex escape on a changed leaf cycle. The right panel
is the implication used in the proof, not a sampled trajectory: a bounded
compensating energy forces $t=O(h)$, contradicting the limiting cycle
equation.}
\label{fig:escape}
\end{figure}

\begin{lemma}[Finite-energy escape]\label{lem:escape}
On one cycle, suppose the fixed target entries are nonzero. Fix generic
positive query factors except at a vertex $p$, whose factor is $h$.
For each fixed finite $R\in\C$, there is $\varepsilon>0$ such that
no root of $G_\ell$ has $E_\ell(t)=R$ when $0<h<\varepsilon$.
\end{lemma}
\begin{proof}
The two incident weighted edge values are $h a$ and $h b$; all other
$z_e$ are independent of $h$. Differentiating \eqref{eq:cycle-polynomial}
at $t=0$ and then setting $h=0$ gives
\begin{equation}
 G_\ell(0,0)=
 \prod_{e\in C_\ell^+\setminus\{e_+(p)\}}z_e+
 \prod_{e\in C_\ell^-\setminus\{e_-(p)\}}z_e=:M_p.\label{eq:M}
\end{equation}
The two terms have nonzero fixed target coefficients and distinct
monomials in the remaining query factors: after removing the incident
edge, the two neighboring vertices occur differently. Thus $M_p$ is not
identically zero. Generic positive outside factors give $M_p\neq0$ even
when the target has mixed signs.

There are real constants $\alpha>0,\beta,\sigma,\tau$, independent of $h$,
such that
\[
 A_\ell=\alpha h^{-2}+\beta,\qquad T_\ell=\sigma h^{-1}+\tau.
\]
If $E_\ell(t)=R$, setting $w=t/h$ yields
\begin{equation}
 (\alpha+\beta h^2)w^2+2(\sigma+\tau h)w-R=0.\label{eq:bounded-roots}
\end{equation}
For small $h$, this quadratic has leading coefficient bounded away from
zero and bounded remaining coefficients. Its two complex roots are
uniformly bounded. If roots of $G_\ell$ with energy $R$ existed along
$h\downarrow0$, then $t=O(h)$ and $G_\ell(h,t)\to M_p\neq0$, a contradiction.
Repeated roots cause no difficulty.
\end{proof}

\begin{lemma}[Coupled nonvanishing specialization]\label{lem:coupled}
Fix a nonempty set $J$ of changed cycles, all with $d_\ell>1$.
There is a positive query, off all relevant leading-coefficient exceptions,
at which no tuple of roots $G_\ell(t_\ell)=0$, $\ell\in J$, has
$\sum_{\ell\in J}E_\ell(t_\ell)=0$.
\end{lemma}
\begin{proof}
The union of the selected cycles is again a cactus. Its block-cut forest
has a leaf cycle with at most one vertex shared with another selected
cycle. Choose a vertex $p$ absent from all other selected cycles. Shared
vertices of cycles outside $J$ do not matter, since those cycles keep
parameter zero and energy zero.

Fix generic positive factors outside $p$ and vary its factor $h$.
Every other selected cycle has finitely many complex roots, independent
of $h$, and hence their combined energies range over a finite set.
For each negative of such an energy, Lemma~\ref{lem:escape} excludes a
compensating leaf root for all sufficiently small positive $h$.
One $h$ works for the whole finite set. The leading coefficients are
nonzero polynomials; generic outside factors and exclusion of finitely
many further values of $h$ keep them nonzero. The query can be placed
inside the unit bounds by initially choosing the outside factors small.
\end{proof}

\subsection{An explicit polynomial certificate}

For degree counting, use $u_i=1/x_i$ and $v_j=1/y_j$; the constants $m,n$
in \eqref{eq:grid-query} only rescale fixed coefficients and cycle
parameters. Let $V_\ell$ be the product of all $x_i,y_j$ on cycle $\ell$.
Writing $F_\ell=\sum_{j=1}^{d_\ell}f_{\ell j}t^j$, set
\begin{equation}
 C_{\ell j}=V_\ell f_{\ell j},\qquad c_\ell=C_{\ell,d_\ell}.
 \label{eq:clear}
\end{equation}
Every nonzero $C_{\ell j}$ is homogeneous of degree $2j$: a term in
the coefficient of $t^j$ chooses $k_\ell-j$ disjoint matching edges,
which remove two distinct vertex factors each from $V_\ell$.
In particular $c_\ell$ is nonzero and has degree exactly $2d_\ell$.

Let $\mathbb K=\R(x_1,\ldots,x_m,y_1,\ldots,y_n)$, with the fixed target entries
included among its constants. For a nonempty subset $J$ of cycles with
$d_\ell>1$, put $e_\ell=d_\ell-1$ and consider
\begin{equation}
 \mathcal A_J=\mathbb K[t_\ell:\ell\in J]/
 \bigl(G_\ell/\operatorname{lc}(G_\ell):\ell\in J\bigr),
 \qquad M_J=\dim_{\mathbb K}\mathcal A_J=\prod_{\ell\in J}e_\ell.
 \label{eq:algebra}
\end{equation}
Let $K_\ell$ denote multiplication by $t_\ell$ in the tensor monomial
basis, and set
\begin{equation}
 H_J=\sum_{\ell\in J}(A_\ell K_\ell^2+2T_\ell K_\ell),\qquad
 a_J=\prod_{\ell\in J}c_\ell^2,\qquad N_J=\det(a_JH_J).
 \label{eq:determinant}
\end{equation}
The companion denominators divide $c_\ell$, so $N_J$ is a polynomial.
A finite-algebra element has zero multiplication determinant precisely
when it is not a unit, equivalently when it vanishes in some complex
residue field after extending scalars. This includes nonreduced algebras.
Thus a false tuple with zero total energy forces $\det H_J=0$ off the
leading-coefficient loci. Lemma~\ref{lem:coupled} shows that this rational
function, and hence $N_J$, is nonzero.

Clearing each matrix entry before taking its determinant introduces
unnecessary powers of leading coefficients. We instead clear the norm
by iterated univariate resultants. Write
$\widetilde G_\ell=V_\ell G_\ell
=\sum_{j=1}^{d_\ell}C_{\ell j}t^{j-1}$ and
$E_\ell(t)=A_\ell t^2+2T_\ell t$. Starting with
$\sum_{\ell\in J}E_\ell(t_\ell)$, successively take its resultant with
$\widetilde G_\ell$ in $t_\ell$, with $\widetilde G_\ell$ the first
argument. The resulting polynomial is
\begin{equation}
 R_J=\left(\prod_{\ell\in J}c_\ell^{\,2M_J/e_\ell}\right)\det H_J.
 \label{eq:resultant-clear}
\end{equation}
Indeed, after a prefix $I$ has been eliminated, each remaining variable
has degree $2M_I$, where $M_I=\prod_{\ell\in I}e_\ell$.
Its leading coefficient is a nonzero product of earlier leading
coefficients and $A_j^{M_I}$. The root-product formula for each
resultant therefore gives \eqref{eq:resultant-clear} by induction.
Roots are counted with multiplicity, so squarefreeness is unnecessary.
In particular $R_J$ is a nonzero polynomial and
$N_J=R_J\prod_{\ell\in J}c_\ell^{2M_J-2M_J/e_\ell}$.

The degree can be counted without expanding a determinant. Under
$(x,y)\mapsto(\lambda x,\lambda y)$, each cycle root scales as
$t_\ell\mapsto\lambda^{-2}t_\ell$, while $A_\ell,T_\ell$ have degrees
four and two. Equivalently, its companion matrix transforms as
$K_\ell\mapsto\lambda^{-2}S_\ell K_\ell S_\ell^{-1}$ with
$S_\ell=\operatorname{diag}(1,\lambda^2,\ldots,\lambda^{2(e_\ell-1)})$.
Consequently $H_J$ transforms by tensor similarity and $\det H_J$ has
degree zero as a rational function. Hence
\begin{equation}
 \deg R_J=4M_J\sum_{\ell\in J}\frac{d_\ell}{e_\ell},\qquad
 \deg N_J=4M_J\sum_{\ell\in J}d_\ell.
 \label{eq:det-degree}
\end{equation}
Both nonzero polynomials are homogeneous. These are the degrees of the specified certificates.

\begin{proof}[Proof of Theorem~\ref{thm:main}]
If $d_\ell=1$, then $G_\ell$ is a nonzero constant off $c_\ell=0$,
so that cycle cannot change. Form the nonzero polynomial
$P_B=(\prod_\ell c_\ell)\prod_{J\neq\varnothing}R_J$, where $J$
ranges over subsets of the $r_+$ cycles with $d_\ell>1$.
It contains every ambiguity. For each eligible cycle,
\begin{equation}
 \sum_{J\ni\ell} M_J\frac{d_\ell}{e_\ell}
 =d_\ell\prod_{j\neq\ell}(1+e_j)=D.
 \label{eq:subset-sums}
\end{equation}
Equations~\eqref{eq:det-degree}--\eqref{eq:subset-sums} give
$\deg P_B=2\sum d_\ell+4r_+D=L_G$.
A nonzero polynomial of degree $L_G$ vanishes at at most an $L_G/N$
fraction of an independent $N$-point grid; Appendix~\ref{app:grid}
gives an elementary proof. Its zero set also has Lebesgue measure zero.
Every actual competitor satisfies the cycle equations used above,
including zero competitors, so excluding this larger set is sufficient.
This proves the $L_G$ estimate. Proposition~\ref{prop:refined-grid}
removes redundant factors and gives the stated $\widetilde L_G$ estimate.
\end{proof}

Appendix~\ref{app:certificate-heights} gives rational coefficient-height bounds for this certificate.

\subsection{A sharper certificate on positive grids}\label{sec:refined-grid}

The certificate can exploit positivity of the query and the fact that a
changed cycle has a nonzero parameter. Put
$\mathcal I=\{\ell:d_\ell>1\}$ and
$\mathcal E_0=\{\ell:k_\ell\text{ is even}\}$.

\begin{proposition}[Refined grid bound]\label{prop:refined-grid}
Under the hypotheses of Theorem~\ref{thm:main}, its failure probability is
at most $\min\{1,\widetilde L_G/N\}$, where
\begin{equation}
 \widetilde L_G=2\sum_{\ell\in\mathcal E_0}d_\ell
                    +4r_+D-2r_+.\label{eq:refined-grid-bound}
\end{equation}
For $s$ six-cycles this gives $4s3^s-2s$; in particular two six-cycles
give $68/N$ in place of $84/N$.
\end{proposition}
\begin{proof}
For $\ell\in\mathcal I$, define the polynomial
\[
 S_\ell=\Res_{t_\ell}\bigl(\widetilde G_\ell,
                                      A_\ell t_\ell+2T_\ell\bigr).
\]
Resultant multiplicativity gives
$R_{\{\ell\}}=C_{\ell1}S_\ell$.
Since $R_{\{\ell\}}$ is nonzero, so is $S_\ell$.
Writing $e_\ell=d_\ell-1$, the linear-resultant formula is
\[
 S_\ell=\sum_{j=0}^{e_\ell}C_{\ell,j+1}
                    (-2T_\ell)^j A_\ell^{e_\ell-j}.
\]
Each term has query degree $4d_\ell-2$, hence so does $S_\ell$.
A false competitor changing just cycle $\ell$ has $t_\ell\ne0$;
its zero energy forces $A_\ell t_\ell+2T_\ell=0$, so it forces
$S_\ell=0$ without using $C_{\ell1}=0$ as an exception.

If $k_\ell$ is odd, then $c_\ell=2V_\ell$, which never vanishes on
the positive reciprocal grid. Retain leading factors only for even
$k_\ell$, and use
\[
 \widetilde P_B=
 \Bigl(\prod_{\ell\in\mathcal E_0}c_\ell\Bigr)
 \Bigl(\prod_{\ell\in\mathcal I}S_\ell\Bigr)
 \prod_{\substack{J\subseteq\mathcal I\\|J|\geq2}}R_J.
\]
Every factor is nonzero. Off its leading factors, a nonempty changed set
has either one cycle, covered by $S_\ell$, or at least two, covered by
the unchanged $R_J$ argument. Thus every ambiguous positive query
vanishes on $\widetilde P_B$. Relative to $P_B$, the total degree drops
by $2\sum_{k_\ell\text{ odd}}d_\ell+2r_+$, proving
\eqref{eq:refined-grid-bound}. The same grid zero-count bound applies;
the fixed rescalings by $m,n$ preserve degree.
\end{proof}

The refined certificate gives the stated grid probability in
Theorem~\ref{thm:main}. Appendix~\ref{app:certificate-heights} bounds the
coefficient heights of the original $P_B$ representation.

\subsection{Finite-grid probabilities and certificate degree}\label{sec:grid-sharpness}

The next two results separate arithmetic collision probability from
the degree of a global exception polynomial. Four-cycle flowers have
an exact grid law that attains the certificate's linear coefficient.
Six-cycle flowers require exponentially large polynomial degree.

\begin{samepage}
\begin{proposition}[Four-cycle flowers]\label{prop:grid-sharpness}
Let $G_r$ consist of $r\geq1$ four-cycles sharing row $0$, with private
row $\ell$ and private columns $(\ell,+),(\ell,-)$ on cycle $\ell$.
Put $\lambda=1/(2r+2)$ and fix the factors
\[
 a=\lambda(1,-1,\ldots,-1),\qquad
 b=\lambda(1,-1,\ldots,1,-1).
\]
Both factor norms are strictly below one and every observed entry is
nonzero. Under the reciprocal grid in Theorem~\ref{thm:main}, the
probability that the complete reply has another actual real candidate is
\begin{equation}
 p_{r,N}=1-(1-1/N)^{2r}.\label{eq:exact-grid-law}
\end{equation}
If competitors must also have unit-bounded factors, its ambiguity
probability is at least $1-(1-1/N)^r$. In particular,
$p_{r,N}\sim 2r/N$ for fixed $r$ as $N\to\infty$, attaining the
first-order coefficient $\widetilde L_{G_r}=2r$ of the general bound.
\end{proposition}
\end{samepage}
\begin{proof}See Appendix~\ref{app:grid-degree-proofs}.\end{proof}

The four-cycle law has a sharp linear coefficient. For six-cycle
flowers the next result shows that exponential polynomial-certificate
degree is unavoidable, while leaving the integer-grid probability
question separate.

\begin{proposition}[Exponential exception-certificate degree]
\label{prop:certificate-degree}
For every $r\geq1$ there is a fixed rational target on a flower of
$r+1$ six-cycles, nonzero on every edge and with strictly unit-bounded
factors, such that every nonzero real polynomial whose zero set contains
all its ambiguous legal positive queries has degree at least $(3^r-1)/2$.
The same lower bound holds in reciprocal query coordinates, already in
one specified coordinate. Ambiguity may be restricted to nonzero
competitors with unit-bounded factors.
\end{proposition}
\begin{proof}See Appendix~\ref{app:grid-degree-proofs}.\end{proof}

The upper coefficient $4(r+1)3^{r+1}-2(r+1)$ is within a linear factor
of this lower bound. Exponential dependence on cycle count is therefore
intrinsic to a global polynomial exception certificate. The integer-grid rate also depends on arithmetic: most constructed
collision coordinates $x_T$ are quadratic irrational numbers. The optimal
six-cycle failure probability on independent integer grids remains open.

\FloatBarrier
\subsection{A normalization witness beyond cactus masks}\label{sec:theta}

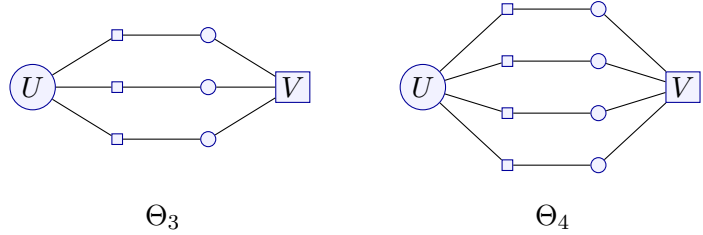
\begin{figure}[htbp]\centering
\begin{tikzpicture}[scale=.86,row/.style={circle,draw=blue!60!black,fill=blue!5,inner sep=2pt},
 col/.style={rectangle,draw=blue!60!black,fill=blue!5,inner sep=2pt}]
\foreach \shift/\p/\name in {0/3/3,6/4/4}{
 \begin{scope}[xshift=\shift cm]
 \node[row] (u\name) at (0,0) {$U$};
 \node[col] (v\name) at (4,0) {$V$};
 \foreach \i in {1,...,\p}{
  \pgfmathsetmacro{\h}{.8*(\i-(\p+1)/2)}
  \node[col] (c\name\i) at (1.3,\h) {};
  \node[row] (r\name\i) at (2.7,\h) {};
  \draw (u\name)--(c\name\i)--(r\name\i)--(v\name);
 }
 \node at (2,-2) {$\Theta_{\p}$};
 \end{scope}
}
\end{tikzpicture}
\caption{Overlapping-cycle blocks covered by Theorem~\ref{thm:theta-blocks}.
Each path has three edges; paths share only the two endpoints. Circles
are row vertices and squares are column vertices. Any pair of paths
forms a six-cycle, so these blocks are not cacti.}
\label{fig:theta-blocks}
\end{figure}

Private-vertex escape uses the edge-disjoint cycle structure of a
cactus. Theta blocks have overlapping six-cycles and require a common
witness instead. An inverse-factor specialization makes every changed
edge contribute a strictly negative real energy change. This proves
Theorem~\ref{thm:theta-blocks} for arbitrary path count and excludes
cancellation when blocks meet at cut vertices.
Figure~\ref{fig:theta-blocks} illustrates two members of the family.

\begin{lemma}[Inverse-factor specialization]\label{lem:normalization}
Suppose every nontrivial block of $G$ is an even cycle or some $\Theta_p$.
Fix an actual real target $B$ nonzero on every cycle edge.
There is a real query $q^*$ with nonzero coordinates at which the complex
cycle-consistent gradient fiber is finite, the true point is simple,
and every false point has
\[
 \operatorname{Re}\bigl(\norm C^2-\norm B^2\bigr)<0.
\]
The squared norms in this assertion are polynomial sums of squares.
The query $q^*$ is a proof specialization and can have negative coordinates.
\end{lemma}
\begin{proof}
Choose finite real factors $B_{ij}=a_i b_j$. We first choose nonzero
positive vertex multipliers $\rho_i,\sigma_j$. On a cycle their edge
products will all equal one. On a $\Theta_p$ block set $\varepsilon=p$;
the desired successive products along every path are
$(\varepsilon,1,\varepsilon)$. Local vertex factors achieving these
products assign $\varepsilon$ to both shared endpoints and one to the
internal vertices. These assignments can be glued along the block--cut
forest: each new block meets the previously assigned vertices at one
cut vertex, whose multiplier is matched by multiplying all its local
row factors by a positive scalar and its column factors by the
reciprocal. This preserves all local edge products. Bridges and isolated
vertices permit arbitrary remaining nonzero multipliers.
Set $u_i^*=\rho_i/a_i$ when $a_i\ne0$, and set it to one otherwise;
define $v_j^*$ similarly. On nontrivial blocks, the true weighted entries
are exactly the prescribed products. Bridge entries are fixed by the
margins. The incidence kernel is the direct sum
of the cycle spaces of the blocks, so each block can be considered
separately, including when blocks share cut vertices.

On a $2k$-cycle its weighted entries become $1+t,1-t,\ldots,1-t$.
The equation is $(1+t)^k=(1-t)^k$. Its finite roots are
\[
 t=\frac{\zeta-1}{\zeta+1},\qquad \zeta^k=1,\quad\zeta\ne-1.
\]
They are simple; every nonzero root is purely imaginary. Undoing the
weights gives $C_e=B_e(1\pm t)$, and therefore every changed edge has
$\operatorname{Re}(C_e^2-B_e^2)=B_e^2t^2<0$.

On $\Theta_p$, write the successive entries on path $i$ as
$(\varepsilon+t_i,1-t_i,\varepsilon+t_i)$.
Margin equality gives $\sum_i t_i=0$.
All its cycle binomials are
\begin{equation}
 (\varepsilon+t_i)^2(1-t_j)
 -(\varepsilon+t_j)^2(1-t_i)=0\quad(i<j).
 \label{eq:theta-unit}
\end{equation}
No solution has $t_i=1$: the equations would force every $t_j=1$,
contradicting the zero sum. Hence $(\varepsilon+t_i)^2/(1-t_i)$ has one common
value $R$, and each $t_i$ is a root of
$t^2+(2\varepsilon+R)t+\varepsilon^2-R=0$.
If all $t_i$ are equal, they are zero. Otherwise
there are two values $\alpha,\beta$, with multiplicities $j,p-j$.
Relabel them so $j\leq p/2$, and set $c=(p-j)/j$ and
$K=\varepsilon^2+2\varepsilon$. Vieta's formulas give
\begin{equation}
 \alpha=-c\beta,\qquad c\beta^2+(1-c)\beta+K=0.
 \label{eq:theta-beta}
\end{equation}
Here $1\leq c\leq p-1<\varepsilon$. The discriminant is negative, and
\[
 \operatorname{Re}\beta=\frac{c-1}{2c},\qquad
 \operatorname{Re}\beta^2=\frac{(c-1)^2-2cK}{2c^2}<0.
\]
The physical outer-edge values are $B_e(1+t_i/\varepsilon)$ and the
middle-edge values are $B_e(1-t_i)$. Their real energy changes have the
signs of $\operatorname{Re}(t_i^2+2\varepsilon t_i)$ and
$\operatorname{Re}(t_i^2-2t_i)$, respectively.
For $t_i=\alpha$ the first is negative, and for $t_i=\beta$ the second
is negative, directly from the signs above. The two remaining cases are
\[
 \operatorname{Re}(\beta^2+2\varepsilon\beta)
 =\frac{(c-1)^2-2c\varepsilon(\varepsilon+3-c)}{2c^2}<0,
 \qquad
 \operatorname{Re}(\alpha^2-2\alpha)
 =\frac{c^2-1-2cK}{2}<0.
\]
Indeed, $(c-1)^2<c\varepsilon$, $\varepsilon+3-c>3$ and
$2cK>c^2$. Thus every changed theta edge has strictly decreased real
energy, for any number of paths. Equations~\eqref{eq:theta-beta} also
show that the specialized fiber is finite.

The true root is simple: cycle derivatives at zero are $2k$, while
linearizing \eqref{eq:theta-unit} gives $K(t_i-t_j)$, of rank $p-1$
on $\sum_i t_i=0$. The product of the block fibers is finite.
Every false tuple changes at least one block, and all changed-edge
energy real parts are negative, proving the assertion.
\end{proof}

\begin{proof}[Proof of Theorem~\ref{thm:theta-blocks}]
Use incidence-kernel coordinates relative to the true weighted array.
Each cycle has one parameter. Each $\Theta_p$ has path parameters
$t_i$ with $\sum_i t_i=0$; their physical entries are affine in these
parameters, with rational query coefficients. All actual competitors
satisfy the block cycle binomials and the total norm equation.

First control escape to infinity near $q^*$ from
Lemma~\ref{lem:normalization}. Cycle leading coefficients are nonzero
there, giving locally uniform root bounds. For a theta block, the top
homogeneous forms of its cubic equations, up to sign, are
\[
 t_i t_j(t_i-t_j),\qquad i<j,\qquad \sum_i t_i=0.
\]
Their only common zero is zero: all nonzero $t_i$ would have to equal
one value, whose positive integer multiplicity times that value is zero.
Thus theta roots are uniformly bounded under small complex query
perturbations. This also proves finiteness of their affine fibers.

Suppose ambiguous complex queries approached $q^*$. Boundedness gives
a convergent subsequence of false parameter tuples. A false limiting
tuple is excluded by the strict real-energy inequality of
Lemma~\ref{lem:normalization}. A true limiting tuple is excluded by the
implicit function theorem: a full-rank subsystem of the block equations
has the single local solution zero, which is present for every query.
Consequently a complex neighborhood of $q^*$ contains no ambiguous query.

After clearing query denominators, the incidence-parameter equations,
norm equality and requirement that some parameter is nonzero define a
constructible set. Its query projection is constructible by Chevalley's
theorem. It cannot be Zariski dense, since a dense constructible set
contains a Zariski open set and meets every complex open neighborhood.
Its proper algebraic closure therefore admits a nonzero vanishing
polynomial. Multiplying that polynomial by its coefficientwise complex
conjugate gives a nonzero real exceptional polynomial. Its zero
set has measure zero in the positive query region. No normalization is
applied to the observer's query or supplied to the decoder.
\end{proof}

\paragraph{Exact block reduction.}
The same incidence elimination gives a constructive finite-list problem.
A theta block contributes $p-1$ variables and $\binom p2$ cubic
equations. The leading-form argument above holds for every affine
margin base, so each such fiber is finite and has at most $3^{p-1}$
complex points by the affine B\'ezout bound. Thus, off the cycle leading
exceptions, at most
\[
 D_{\rm blocks}=\Bigl(\prod_{\text{cycle blocks }\ell}d_\ell\Bigr)
               \prod_{\text{theta blocks }\Theta_p}3^{p-1}
\]
tuples enter the global norm and support tests. Exact real solutions of
the block systems can be represented by rational univariate
representations \citep{rouillier1999}. Joining their energy lists and
applying Lemma~\ref{lem:actual} then gives the complete actual list.
Appendix~\ref{app:rur} gives an executable quotient-algebra construction
and a completeness certificate for algebraic candidates, including
nonradical finite reply ideals.
The explicit $\widetilde L_G/N$ bound and coefficient-height formula
proved above continue to concern the cactus subclass.

\paragraph{Certificates for the two block classes.}
The cactus construction gives explicit degree, coefficient-height and
finite-grid bounds. For theta blocks, Chevalley's theorem turns the
explicit specialization into an existence proof for the exceptional
polynomial; $3^{p-1}$ bounds candidate points. Extending the fixed-target
guarantee to arbitrary bipartite masks, or deriving theta degree and
finite-grid bounds, remains open.

\section{Regularity of the target}\label{sec:regular}

The cactus argument extends from nonzero cycle entries to smooth points
of the observed rank-one variety. A singular two-block construction
then shows how persistent energy exchange can obstruct identification.

\subsection{Smooth targets with zero cycle entries}

Smoothness gives the geometric condition behind private-vertex escape. Let $X_E$ be the complex Zariski closure of the rank-one observed
model. Define $h_\ell(B)=\prod_{C_\ell^+}B_e-\prod_{C_\ell^-}B_e$.

\begin{lemma}\label{lem:smooth}
For a cactus, $X_E$ is the product of the cycle hypersurfaces $h_\ell=0$
and free bridge coordinates. A point is smooth precisely when
$\nabla h_\ell\neq0$ for every cycle. If some cycle entries are zero,
this condition is equivalent to one alternating matching having exactly
one zero entry.
\end{lemma}
\begin{proof}
The cycle equations use disjoint edge-coordinate sets. Each binomial is
irreducible, so their product variety is irreducible of dimension
$|E|-r$. The toric image is contained in it and has the same dimension:
the exponent matrix is the incidence matrix, of rank $|E|-r$ by
Lemma~\ref{lem:incidence}. Thus the two closures agree.
The Jacobian criterion reduces smoothness to each separate gradient.
If some cycle entries vanish, both matching products vanish. A partial
derivative is nonzero precisely when the product remaining after removal
of one matching entry is nonzero, which is the stated criterion.
\end{proof}

\begin{proposition}[Regular-target extension]\label{prop:regular}
The conclusions and the same bound of Theorem~\ref{thm:main} hold for
every fixed actual real target $B$ that is smooth on $X_E$.
\end{proposition}
\begin{proof}
For a selected leaf cycle containing zeros, choose a matching with a
unique zero edge $e$. At most one cycle vertex is shared with another
selected cycle, so at least one endpoint $p$ of $e$ is private. In
\eqref{eq:M}, the term for this matching removes its only zero entry
and is nonzero. The other term may vanish; if both are nonzero their
distinct query monomials prevent identical cancellation. The two target
entries incident to $p$ need not themselves be nonzero. Nevertheless
$A_\ell=\alpha h^{-2}+\beta$ with $\alpha>0$ still holds, and
\eqref{eq:bounded-roots} gives the same contradiction.
Regularity also ensures some cycle entry is nonzero, so
$\sum_{e\in C_\ell}z_e$ remains a nonzero query polynomial for even
$k_\ell$. All finite-algebra degree bounds are unchanged.
\end{proof}

The relevant smoothness is that of the image closure. Some singular
targets are also identifiable: the zero target is determined by its
zero norm. The following construction exhibits a different singular
stratum, where an open set of complete replies remains ambiguous.

\subsection{A persistent singular ambiguity}\label{sec:boundary}

\begin{proposition}[Singular cactus boundary]\label{prop:singular}
There is a connected bipartite cactus with two cycles and a fixed
$B\in\Bobs^{(1)}$ with zero entries for which an open set of legal
positive queries has distinct actual bounded targets with identical
complete replies.
\end{proposition}
The construction in Appendix~\ref{app:singular} moves energy from a
six-cycle to a previously zero four-cycle. This persistent exchange
identifies the singular obstruction excluded by the regular-target result.

\section{Computational validation}\label{sec:verification}

The computations test three parts of the reconstruction argument:
coverage of the complete actual list, singleton certification before
combination, and the cost of exact energy matching. Finite-grid studies
then compare the information channels and the exception bounds.
The primary implementation uses SymPy~\citep{sympy2017}. A separate
observed-edge path uses Singular~\citep{greuel2008} for quotient and
radical computations, with separate standard-library rational and Sturm
checks. The supplement specifies the input cohorts, timing boundaries
and complete outcomes.

\subsection{Independent coverage of the complete real list}
The observed-edge implementation completes all $37$ prelisted
coverage jobs on $34$ distinct public inputs: $27$ singleton, seven
multiple and three empty actual lists. These include irrational
candidates, nonradical reply ideals and optional factor budgets.
For $35$ nonempty ambient ideals, exact modular composition identities
identify the two independent RURs; Sturm intervals match every real
embedding and its support decision. Two remaining controls check
inconsistent margins and nonactual support directly. All $41$ returned
candidate occurrences agree, including ten nonrational occurrences.
Three corrupted certificates---a missing root, a changed coordinate and
an altered support decision---are rejected. The independent path uses
Singular for Gr\"obner and radical computations and separate rational
and Sturm code for the checks; it imports neither SymPy nor the primary
incidence reduction.

\subsection{Singleton certificates before energy matching}
The energy-coordinate preprocessor of Proposition~\ref{prop:squareclass}
uses only the public mask, query and reply. On all $1605$ six-cycle
queries, it certifies $1486$ singletons, compared with $708$ under
private-class deletion alone. Every returned array passes exact support
checks and reproduces both gradients and the loss. Of the certificates,
$125$ already have singleton gradient fibers; $1361$ require branch
deletion, with up to $531441$ actual gradient candidates before pruning.
All decisions agree with the complete-fiber counts. The remaining
$119$ replies contain $54$ singletons and all $65$ ambiguous replies.
Table~\ref{tab:squareclass} records the resulting certificate frequency
on this fixed cohort. The largest observed number
of deletion passes is five.

\begin{table}[htb]\centering
\caption{Singletons certified before a product join on the
six-cycle cohort. Each cell has $64$ specified queries.
At $N=1$, the one-cycle reply is certified and the other four remain unresolved.}
\label{tab:squareclass}
\begin{tabular}{rrrrrr}\toprule
$r$ & $N=2$ & $4$ & $8$ & $16$ & $32$\\\midrule
1 & 59 & 64 & 64 & 64 & 64\\
2 & 60 & 64 & 64 & 64 & 64\\
4 & 54 & 64 & 64 & 64 & 64\\
8 & 33 & 57 & 63 & 64 & 64\\
12 & 19 & 52 & 64 & 64 & 64\\\bottomrule
\end{tabular}
\end{table}

A separate $96$-input protocol fixes new rational signed targets and
queries before execution: $r\in\{1,2,4,8,12,24\}$,
$N\in\{2,4,8,32\}$, four inputs per pair. It certifies $84$ singletons
and leaves $12$ replies unresolved. All $48$ inputs with $r\leq4$
also receive complete tuple checks. A certified $24$-cycle input has
$31381059609$ real root tuples before propagation; the preprocessor
joins none of them.

The rational subset-sum controls give five singleton certificates,
nine empty certificates and $18$ unresolved inputs out of $32$.
On the $14$ uniform-protocol inputs it certifies eight singletons and
leaves six unresolved, including four higher-degree cases.
Each deletion certificate is checked separately; a rational Sturm chain
checks root-list coverage without repeating the generator's factorization.
Additional controls comprise $500$ finite-domain exhaustive oracles,
six Sturm cases, three radical-coordinate cases, two empty masks and
four deliberately corrupted certificates, which are rejected. The generator
and checker share rational arithmetic and incidence reduction.

\subsection{Exact energy matching and measured scaling}
All three methods run on the same $14$ public inputs with three
serial repetitions, fixed CPU affinity and a $30$-second after-ready
limit on Linux/WSL (Ryzen 7 5800U). The interval includes input reading,
equation construction, exact root work, native startup, support tests
and output serialization. Parent imports are recorded separately.
The cycle join completes all $42$ runs, msolve completes $33$ and
times out in nine, and observed-edge Singular completes $27$ and
times out in fifteen. Every completed list agrees exactly with its
reference. The msolve route uses the reduced cycle system
\citep{berthomieu2021}; Singular starts from observed-edge equations.
The cycle representation completes every run in this cohort, with
the larger flowers separating its completion pattern from the general
solver routes. The comparison measures complete implementations and
representations on these synthetic inputs. Supplement S3 reports
per-input medians, including cases where msolve is faster, and all
censored runs.

The operation counts expose the combination mechanism. An eight-cycle
input has $140625$ real root tuples, but its join uses two lists of $375$
and $5488$ exact comparisons, with one energy match. The largest
twelve-cycle input has $6834375$ real tuples and partial lists of $3375$
and $2025$. The partial-list sizes expose how the join avoids visiting the full
Cartesian product on these inputs.

\begin{figure}[tb]\centering
\includegraphics[width=\linewidth]{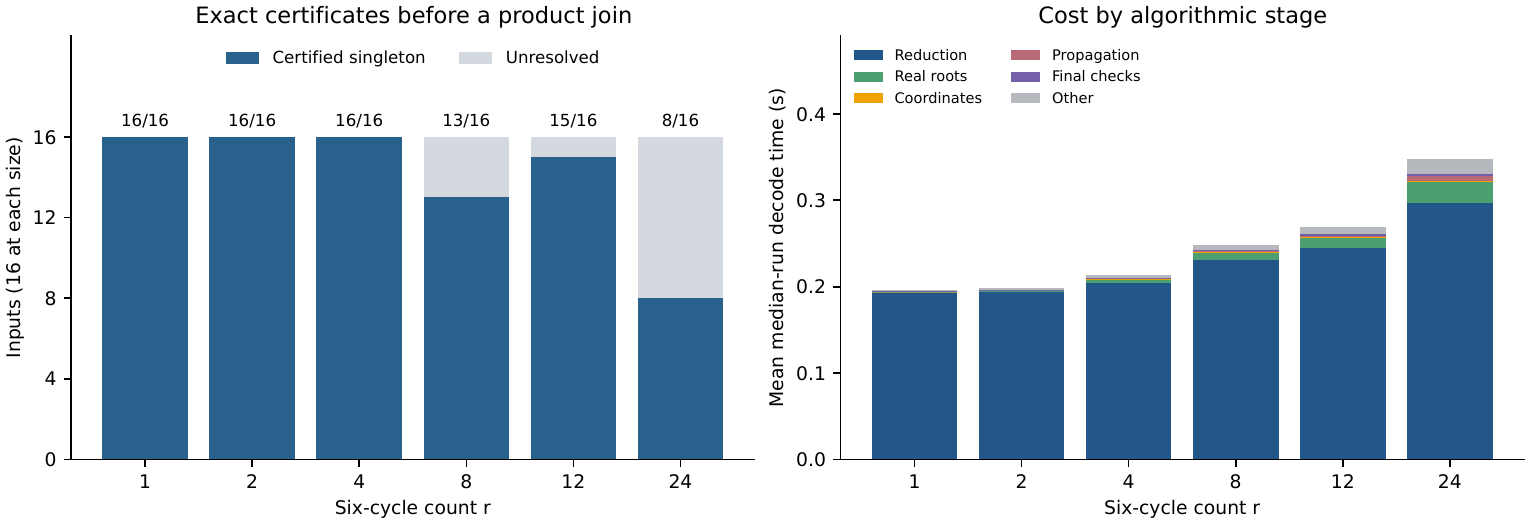}
\caption{Energy-coordinate certificates and their measured cost on all
$96$ prospective synthetic inputs. Left: certificates among the sixteen
inputs at each cycle count; unresolved replies remain included. Right:
phase times for each input's median-total run, averaged by size. The
phase protocol excludes imports and public JSON parsing. These are
certificate frequencies and single-host timings. All $288$ repeated
calls reproduce the exact saved results without joining root tuples.}
\label{fig:mechanism-scaling}\end{figure}

The coordinate preprocessor is measured separately on all $96$ inputs,
three times each. All $288$ runs reproduce the saved outcomes exactly:
$84$ singleton certificates and $12$ unresolved replies per repetition.
Figure~\ref{fig:mechanism-scaling} separates reduction, roots, coordinates,
propagation and final checks. No Cartesian tuples are joined in these runs. Public preprocessing
accounts for most measured time at the tested sizes, identifying where
the computational work remains after branch elimination.

\subsection{Finite-grid laws and certificate degree}
The grid studies test the loss's separating effect and the arithmetic
of exceptional queries. Exhaustion of all $5615$ points on ten
four-cycle grids verifies the exact law; all $3675$ collision witnesses
preserve both gradients and total norm. The six-cycle cohort contains
$1605$ queries and $65$ complete-reply collisions, with one collision
among the $1280$ queries at $N\geq4$. All draws and repetitions are
retained. The sufficient six-cycle bound is vacuous on the tested grids
for $r\geq2$, so these frequencies describe the specified discrete laws.
Separately, $1821$ distinct algebraic slice witnesses check the
certificate-degree construction. The supplement gives the exhaustive
grids, sampled cohorts and exact protocols.

\FloatBarrier
\begin{samepage}
\section{Conclusion}
The scalar loss supplies the information that weighted margins can leave
unresolved. On cycle and length-three parallel-path blocks it identifies
every fixed target with nonzero cycle entries almost surely, even on
open families with exponentially many bounded gradient candidates.
Private-vertex escape and a common inverse-factor witness exclude
cross-block energy cancellation, yielding explicit cactus certificates
and the theta identification guarantee. Exact root lists, energy matching
and support tests turn the cactus result into certified reconstruction;
energy-coordinate propagation can establish uniqueness before combining
the lists. Together with the degree, output and fixed-query hardness
results, these give a precise account of both the information supplied
by one loss and the arithmetic required to recover it.
\par
\end{samepage}

\appendix
\section{Polynomial grid counting}\label{app:grid}

\begin{lemma}
If a nonzero polynomial in $q$ variables has total degree at most $d$,
and each variable is sampled independently uniformly from a set of $N$
distinct real numbers, its zero probability is at most $d/N$.
\end{lemma}
\begin{proof}
Induct on $q$. Write $P=\sum_{j=0}^a P_j(x_1,\ldots,x_{q-1})x_q^j$
with $P_a\neq0$. The induction hypothesis bounds $P_a=0$ by
$(d-a)/N$. Otherwise the univariate polynomial has at most $a$ roots,
so the remaining conditional probability is at most $a/N$.
The sum gives $d/N$. The continuous zero-set statement follows by the
same induction and Fubini's theorem.
\end{proof}

\section{An open singular-target ambiguity on a cactus}\label{app:singular}

Use five rows, six columns and edges
\begin{align*}
 C_6&:(00,01,11,12,22,20),\\
 C_4&:(33,34,44,43),\qquad\text{bridges}:(25,35).
\end{align*}
The graph is connected with cycle rank two. Before division by sixteen,
the true and false factors at the uniform query are
\begin{align*}
 a&=(0,1,-3/2,0,0),& b&=(1,1,-1,1,-1,0),\\
 a'&=(-1,0,-1/2,1,-1),& b'&=(1,-1,-1,1,-1,0).
\end{align*}
Divide every factor by four, so both observed targets are divided by
sixteen. Their left and right squared factor norms are $13/64$ and
$5/16$. With $u_i=1/5$, $v_j=1/6$, their complete replies coincide.
The raw true $C_6$ entries are $(0,0,1,-1,3/2,-3/2)$, of energy $13/2$;
the alternate entries are $(-1,1,0,0,1/2,-1/2)$, of energy $5/2$.
The formerly zero $C_4$ gains the alternating rank-one array of energy four.

\begin{proof}[Proof of Proposition~\ref{prop:singular}]
Analyze positive raw queries $U,V$ near all ones; multiplying them by
$1/5,1/6$ preserves candidate margin equalities. On $C_6$, the first
two weighted entries are zero and
\[
 F(t)=tg_{\rm aux}(t),\qquad
 g_{\rm aux}(t)=2t^2+(z_2+z_4-z_3-z_5)t+z_2z_4+z_3z_5.
\]
At the base, $g_{\rm aux}(t)=2(t+1)(t+3/2)$. The root $-1$ is simple, so a real
root $t(U,V)$ near $-1$ persists on a full query neighborhood. The
candidate $C_e=(z_e+t n_e)/(U_i V_j)$ has the same margins. Its nonzero
row-0 and row-2 entries allow actual factor propagation with
\[
 b_0=1,\ a_0=C_{00},\ a_2=C_{20},\ b_1=C_{01}/a_0,
 \ b_2=C_{22}/a_2,\ a_1=C_{11}/b_1.
\]
The last edge equation follows from $F(t)=0$. These factors remain
continuous even if row 1 becomes zero.

The $C_6$ energy decrease $\Delta$ equals four at the base and stays
positive. On the zero $C_4$, the gradient kernel array has entries
\[
 N_{\rm ker}=(1/(U_3V_3),-1/(U_3V_4),1/(U_4V_4),-1/(U_4V_3)).
\]
It is rank one, with factors $(1/U_3,-1/U_4)$ and
$(1/V_3,-1/V_4)$. Multiplying it by
$\sqrt{\Delta/\norm{N_{\rm ker}}^2}$ restores exactly the lost energy. Keep the
bridge column's factor zero. The two component factorizations then
coexist in one actual global factorization. After the same normalization,
the strict query and target bounds persist on a smaller open neighborhood.
Margins and total norm agree there, hence so does the complete reply.
The fixed target is singular because the four-cycle binomial has zero
gradient at its zero block.
\end{proof}

\section{Matched information separation and branch conditioning}\label{app:matched-information}

We now compare loss and gradients without changing the target or query
experiment. The example uses two six-cycles sharing a row vertex. On each cycle order the edges as
$(00,01,11,12,22,20)$, using local row and column indices.

Let $\lambda=1/64$ and use the fixed target $B=\lambda(P,M)$, where
\begin{equation}
 P=(9,3,2,-2,-2,6),\qquad M=(9,3,3,-3,-1,3).\label{eq:PM}
\end{equation}
Its local raw factors are $(1,2/3,2/3),(9,3,-3)$ on $P$, and
$(1,1,1/3),(9,3,-3)$ on $M$. Glue the first row factor, then divide
the global left factors by four and the right factors by sixteen.
Use $v_j=1/6$, $u_i=1/5$ except $u_1=h/5$, with $9/10<h<11/10$.
These query norms are strictly below one.

\begin{samepage}
\begin{proposition}[Matched comparison]\label{prop:utility}
For this same fixed $B$, there is a full-dimensional open neighborhood
$U$ of legal positive queries on which gradients alone have at least
four distinct actual nonzero candidates in $\Bobs^{(1)}$. The complete
reply is unique among all actual real rank-one targets for almost every
query in $U$. Along the displayed path, it is unique for the whole
interval $9/10<h<11/10$, except at $h=1$, where a complete-reply collision
remains.
\end{proposition}
\end{samepage}
\begin{proof}
Write $w_e=30u_i v_j$, equal to $h$ on the two row-1 edges of $P$ and one
elsewhere. Candidate cycle arrays have the form
\[
 C_P(t)=\lambda(P_e+t n_e/w_e),\qquad
 C_M(s)=\lambda(M_e+s n_e),\qquad n=(+1,-1,+1,-1,+1,-1).
\]
Direct expansion gives, up to nonzero scale,
\begin{equation}
 F_P(t)=2t(t-1)(t+2h),\qquad F_M(s)=2s(s+1)(s+3).
 \label{eq:PM-roots}
\end{equation}
Thus every actual competitor is one of nine cycle tuples. The norm
increments, divided by $\lambda^2$, are
\begin{equation}
\begin{array}{c|ccc}
 t&0&1&-2h\\\hline
 E_P(t)&0&2/h^2+8/h&16h^2+8h-8
\end{array}
\qquad
\begin{array}{c|ccc}
 s&0&-1&-3\\\hline
 E_M(s)&0&-10&6.
\end{array}\label{eq:energy-table}
\end{equation}
On the interval, the two false $P$ increments are positive and
\begin{align}
 E_P(1)-10&=2(1-h)(5h+1)/h^2,\label{eq:gap}\\
 E_P(-2h)-10&>54/25>0.\nonumber
\end{align}
The second inequality follows from strict monotonicity and its value
$54/25$ at $h=9/10$. Hence only $(t,s)=(0,0)$ has zero total increment
when $h\neq1$. At $h=1$, $(1,-1)$ also has zero increment.

The four tuples $t\in\{0,1\}$, $s\in\{0,-1\}$ are actual and nonzero.
For $t=1$, raw factors on $P$ are
$(1,1+1/(2h),1/2)$ and $(10,2,-2)$. For $s=-1$, those on $M$ are
$(1,1/2,1/2)$ and $(8,4,-4)$. They glue at the shared first row.
With the same four/sixteen normalization, all four choices satisfy
\begin{equation}
 \norm a^2\leq1549/5184<1,\qquad \norm b^2\leq207/256<1.
 \label{eq:utility-caps}
\end{equation}
Every edge has magnitude at least $\lambda$.

At the base query $h=1$, the chosen roots $0,1$ of $P$ have derivatives
$-4,6$, and $0,-1$ of $M$ have derivatives $6,-4$. The implicit-function
theorem continues them under changes of \emph{all} positive query
coordinates. Nonzero cycle solutions factor by rational propagation,
with the first row factor fixed to one. Distinctness, actual factors and
the strict norm bounds persist on a smaller full-dimensional open set.
Finally $B$ satisfies Theorem~\ref{thm:main}, so complete-reply uniqueness
holds almost everywhere in that same set.
\end{proof}

\begin{theorem}[Separation on an open target set]\label{thm:open-separation}
On the two-cycle mask of Proposition~\ref{prop:utility}, there exist a
nonempty relatively open set $\mathcal U$ of all-edge-nonzero actual
targets and a nonempty open set $\mathcal V$ of legal positive queries
with the following properties. Every $B\in\mathcal U$ has factors of
norm strictly below one. For every $(B,q)\in\mathcal U\times\mathcal V$,
the gradient reply admits at least four distinct actual targets with
factor norms strictly below one. For each fixed $B\in\mathcal U$, its
complete reply uniquely determines $B$ among all actual targets for
Lebesgue-almost every $q\in\mathcal V$.
Consequently, under any absolutely continuous query law $\mu$ supported
on legal positive queries with $\mu(\mathcal V)>0$, gradient ambiguity
has probability at least $\mu(\mathcal V)$ for every $B\in\mathcal U$,
whereas complete-reply ambiguity has probability zero.
\end{theorem}
\begin{proof}
Let $B_0$ be the target in \eqref{eq:PM}, and let $q_0$ be its uniform
query ($h=1$). For a target $B$ and a positive query $q=(u,v)$, write
the candidates on each cycle as
\[
 C_e=B_e+\lambda t n_e/(30u_i v_j).
\]
The alternating products of the $C_e$ give a cycle polynomial whose
coefficients depend smoothly on $(B,q)$. At $(B_0,q_0)$ the selected
roots are $t=0,1$ on $P$ and $t=0,-1$ on $M$, with the nonzero
derivatives recorded in Proposition~\ref{prop:utility}. The joint
implicit-function theorem continues all four real simple roots in a
neighborhood of $(B_0,q_0)$. Shrink this neighborhood so that the two
roots on each cycle stay distinct and every candidate edge stays nonzero.

Each pair of selected roots therefore gives a nonzero array with the
same weighted margins as $B$. Fix the raw factor on the shared row to
one and propagate factors along a spanning tree. These factors are
rational functions of the nonzero edge entries. The two cycle equations
ensure consistency at the removed edges, so the propagated factors give
actual rank-one arrays and depend continuously on $(B,q)$. Divide the
left factors by four and the right factors by sixteen, using the raw
array $C/\lambda$ in this propagation. At $(B_0,q_0)$ the strict bounds
\eqref{eq:utility-caps} hold for all four root pairs; they consequently
persist in a smaller neighborhood.

Restrict the target variable to the all-edge-nonzero actual target
space. The neighborhood just obtained contains a product
$\mathcal U\times\mathcal V$ in its relative product topology. The
unchanged branch is the target itself, since its cycle parameter is
always zero. Thus the four candidates, including the true target, all
satisfy the asserted strict factor bounds. Every $B\in\mathcal U$
satisfies Theorem~\ref{thm:main}, which gives complete-reply uniqueness
outside a null set for that fixed $B$. Applying an absolutely continuous
law to this statement and to the common query set $\mathcal V$ gives
the probability assertions.
\end{proof}

\paragraph{Exponential family.}
\begin{proof}[Proof of Corollary~\ref{cor:exponential-separation}]
Use the flower, scale $\lambda=1/(256r)$ and uniform query
$q_0=1/(4r)$ from Proposition~\ref{prop:output-size}, taking its true
target to have parameter zero on every cycle. At this base point each
cycle polynomial is $2\lambda^3s(s^2-3)$. Its roots
$0,\sqrt3,-\sqrt3$ are simple, with derivatives
$-6\lambda^3,12\lambda^3,12\lambda^3$.

For varying $(B,u,v)$, use the affine cycle coordinate
$C_e=B_e+\lambda q_0^2s n_e/(u_i v_j)$ on each cycle.
The joint implicit-function theorem continues all three real roots;
they remain distinct, and all corresponding edge entries remain nonzero.
The leading cubic coefficients also remain nonzero. Thus these are all
the cycle roots, and Lemma~\ref{lem:incidence} gives exactly $3^r$
gradient-compatible arrays.

At the base point, Proposition~\ref{prop:output-size} supplies finite
factors for every branch, with squared norms below $213/256$ and
$146/256$. Fix the shared row factor to $\sqrt\lambda$ and propagate
along a spanning tree. The factors depend continuously on the nonzero
entries; the cycle equations ensure consistency. Since there are
finitely many branches for each fixed $r$, their actualness and strict
factor bounds persist in a common neighborhood. Restricting to actual
targets and taking a product neighborhood gives
$\mathcal U_r\times\mathcal V_r$. Theorem~\ref{thm:main} gives the
fixed-target almost-sure assertion.
\end{proof}

For each fixed target in this open family, the loss almost surely
selects one array from an exponentially large gradient list on the common
query set. The large complete lists of Proposition~\ref{prop:output-size} occur
at exceptional replies. The open neighborhoods in this continuation
argument may depend on $r$.

\begin{remark}[Identifiability and conditioning]
The comparison concerns the candidate fiber at the same query;
alternative targets may depend on that query. Its numerical consequence
is explicit: as $h\to1$, the bounded competitor $(1,-1)$ stays separated
from $B$, while its norm gap tends to zero by \eqref{eq:gap} and its loss
gap is half that value. Thus almost-sure identification can coexist with
arbitrarily small branch-separation margins.
\end{remark}

For the bounded competitor $C_h$ with cycle parameters $(1,-1)$,
the exact branch separation is
\begin{equation}
 \norm{C_h-B}^2=\lambda^2(10+2/h^2),\qquad
 |f_{C_h}-f_B|=\lambda^2\frac{|1-h|(5h+1)}{h^2}.
 \label{eq:conditioning-gap}
\end{equation}
Its gradients equal those of $B$. The secant inverse ratio is asymptotic to
$(\sqrt3\lambda|h-1|)^{-1}$ near $h=1$; any uniform inverse bound
over the family must dominate this ratio. This is a loss of global
branch separation, distinct from local tangent conditioning.

\section{Grid laws and certificate-degree lower bounds}\label{app:grid-degree-proofs}

\begin{proof}[Proof of Proposition~\ref{prop:grid-sharpness}]
Write $c=\lambda^2$. A cycle's physical matrix is
$c(1,-1)^T(1,-1)$. Its total weighted sum is
$c(u_0-u_\ell)(v_{\ell,+}-v_{\ell,-})$. For a four-cycle,
the cycle polynomial is linear with exactly this leading coefficient.
If all these coefficients are nonzero, the gradient fiber is a
singleton. Thus ambiguity requires one of the $2r$ equalities
$x_\ell=x_0$ or $y_{\ell,+}=y_{\ell,-}$.

If $u_0=u_\ell$, reflect the column factor $(\lambda,-\lambda)$
in the line spanned by $(v_{\ell,+},v_{\ell,-})$. This orthogonal
reflection preserves its norm and weighted sum. The weighted row-factor
sum is zero, so both gradient blocks and the cycle energy stay fixed.
The reflection changes the column factor because the query is positive.
Keep every row factor and every other column factor unchanged. This
gives a distinct actual candidate with both global factor norms unchanged.

If $v_{\ell,+}=v_{\ell,-}$, instead reflect the row vector
$(\lambda,-\lambda)$ in the line spanned by $(u_0,u_\ell)$.
The resulting local rank-one array again has the same weighted margins
and norm. Its shared-row factor cannot vanish on this rational grid:
vanishing would require $u_0/u_\ell=1+\sqrt2$. Hence rescale its local
factors to restore shared-row factor $\lambda$, gluing it to all other
cycles. This gives a finite actual competitor; a global unit-factor
bound is not imposed in this case.

Conditionally on $x_0$, the $r$ private-row equality events and the $r$
column-pair equality events are independent, each with probability
$1/N$. Their union gives \eqref{eq:exact-grid-law}; the row events
alone give the bounded-class lower bound. Here all $d_\ell=1$, so
$\widetilde L_{G_r}=L_{G_r}=2r$.
\end{proof}

\begin{proof}[Proof of Proposition~\ref{prop:certificate-degree}]
Use one cycle of type $P$ and $r$ cycles of type $M$, sharing row zero.
Order each cycle as $(00,01,11,12,22,20)$ and use six-edge arrays
\[
 P=(9,3,2,-2,-2,6),\qquad M=(9,3,3,-3,-1,3).
\]
On $M_i$, multiply the array by $\omega_i=2^{i-1}$. Their raw local
factors are $(1,2/3,2/3),(9,3,-3)$ for $P$ and
$(1,1,1/3),\omega_i(9,3,-3)$ for $M_i$. Glue their first row factors,
and divide both global factor vectors by $\gamma=2^{r+4}$.
Put $\lambda=\gamma^{-2}$. This defines a single rational target.
There are $m=2r+3$ rows and $n=3r+3$ columns.

Take $v_j=1/n$ and $u_i=1/m$, except at the first private row of $P$,
where $u_p=1/(mx)$. On each $M_i$ choose one of three branches, with
raw factors and energy change divided by $\lambda^2$ given by
\[
\begin{array}{c|c|c|c}
 \text{branch}&\text{row factors}&\text{column factors}&\Delta_i/\lambda^2\\\hline
 0&(1,1,1/3)&\omega_i(9,3,-3)&0\\
 1&(1,1/2,1/2)&\omega_i(8,4,-4)&-10\omega_i^2\\
 2&(1,0,1)&\omega_i(6,6,-4)&6\omega_i^2
\end{array}
\]
Let $d_i=0,5,-3$ for these choices, and put
\[
 T=\sum_{i=1}^r d_i4^{i-1}.
\]
All $3^r$ sums are distinct. To see this, digit differences lie in
$\{0,\pm3,\pm5,\pm8\}$. At their lowest nonzero position an odd
difference cannot cancel modulo four. A difference $\pm8$ produces a
carry $\pm2$, which no next digit difference cancels modulo four.
The highest nonzero digit determines the sign of $T$. There are thus
$\sum_{k=0}^{r-1}3^k=(3^r-1)/2$ positive sums, all at least five.

For each $T>0$, change the $P$ factors to
$(1,1+x/2,1/2),(10,2,-2)$ and use the chosen $M$ branches, preserving
the global $\gamma$ normalization. The shared row factor remains one.
These arrays have the same weighted margins as the fixed target.
Their total squared-norm change is
\begin{equation}
 \lambda^2\bigl(2x^2+8x-2T\bigr),
 \label{eq:degree-collisions}
\end{equation}
so their complete replies agree at $x_T=\sqrt{4+T}-2$. These are
$(3^r-1)/2$ different collision parameters satisfying
$1\leq x_T<2^{r+1}$. All target and competitor factors
have squared left norm below $8\,4^r/\gamma^2=1/32$ and squared
right norm below $52\,4^r/\gamma^2=13/64$. Indeed, put
$W_{\rm all}=(4^r-1)/3$. We have $0<T\leq5W_{\rm all}$ and
$x_T^2+4x_T=T$. Before normalization, the competitor's left squared
norm is at most $5/4+(1+x_T/2)^2+10r/9<8\,4^r$, and its right squared
norm is at most $108+99W_{\rm all}<52\,4^r$. The target satisfies the
same bounds. The queries also have strict unit-factor bounds.

It remains to exclude a polynomial that vanishes identically on this
particular slice, and to obtain nonzero competitors. Let $\eta$ denote all other reciprocal query
coordinates, initially one. In normalized cycle-flow coordinates the
selected $P$ root is $t=1$ of $2t(t-1)(t+2/x)$, and a selected $M_i$
root is $t=-\omega_i$ or $-3\omega_i$ of
$2t(t+\omega_i)(t+3\omega_i)$; the unchanged roots are zero.
All these roots are simple. Their real actual branches
therefore continue under arbitrary small changes of $(x,\eta)$.
On each $M$ block, the edges $00,01,20,22$ stay nonzero, so factor
propagation divides only by nonzero quantities even on branch 2.
The $x$ derivative of \eqref{eq:degree-collisions} at $x_T$ is
$\lambda^2(4x_T+8)>0$. The implicit-function theorem gives, on a
common open neighborhood of the outside coordinates, $(3^r-1)/2$
distinct collision parameters $x_T(\eta)$. Choose disjoint small
intervals around the original $x_T$; strict factor bounds and legality
persist on this common neighborhood.

Branch 2 initially has one zero row. Restrict the outside neighborhood
to queries with $v_{i,1}\ne v_{i,2}$ on every $M_i$. The fixed target's
weighted margin at that private row is
$3\lambda\omega_i(v_{i,1}-v_{i,2})\ne0$. Equality of margins forces its
continued candidate row factor to be nonzero. The adjacent column
factors stay nonzero, and all other edges were nonzero at the base.
Thus every continued competitor is nonzero on this nonempty open set.

Expand any proposed reciprocal-coordinate certificate as
$Q(x,\eta)=\sum_j q_j(\eta)x^j$. Since $Q$ is nonzero, some
coefficient $q_j$ is a nonzero polynomial. Choose $\eta$ in that open
neighborhood where $q_j(\eta)\ne0$. Then $Q(\cdot,\eta)$ is nonzero
and has at least $(3^r-1)/2$ distinct real roots, proving the partial-degree
bound. Replacing $x$ by the coordinate $u_p=1/(mx)$ gives the same
argument for polynomials in the original query coordinates.
\end{proof}

\section{Integer-lattice uniqueness}\label{app:lattice}
\subsection{Exact uniqueness on an integer energy lattice}\label{sec:lattice}

Exact uniqueness can be decided without printing every matching tuple
when the local energies lie on a small rational lattice. The following
statement isolates this regime from the general algebraic join.

\begin{samepage}
\begin{proposition}[Lattice uniqueness test]\label{prop:lattice}
Suppose nonempty finite lists of distinct local branches have rational energies
$e_{\ell j}$ and every combination is an actual candidate. The unit-factor
bound is omitted. Let $Q_{\rm lat}>0$ clear their denominators and that of the
required cycle energy $\nu$. Put
\[
 c_{\ell j}=Q_{\rm lat}(e_{\ell j}-\min_j e_{\ell j})\in\mathbb Z_{\geq0},
 \qquad U=Q_{\rm lat}(\nu-\sum_\ell\min_j e_{\ell j})\in\mathbb Z.
\]
For $r\geq1$ and $U\geq0$, emptiness, uniqueness and multiplicity can be decided in
$O((U+1)\sum_\ell n_{\ell,\mathbb R})$ operations on counts in $\{0,1,2\}$,
where $n_{\ell,\mathbb R}$ is the number of local branches on cycle $\ell$ and
where $2$ means at least two. A singleton is reconstructed with
$O(r(U+1))$ stored counts and predecessor indices. For $U<0$ the list
is empty. For $r=0$, there is one candidate exactly when $U=0$.
The bound is pseudopolynomial in the integer energy range.
\end{proposition}
\end{samepage}
\begin{proof}
Set $C_0(0)=1$ and other entries to zero. For $0\leq s\leq U$, compute
\[
 C_\ell(s)=\min\{2,\sum_j C_{\ell-1}(s-c_{\ell j})\},
\]
with entries outside $[0,U]$ zero. Induction shows that each entry is
the number of branch prefixes at that energy, capped at two. Capping
preserves both zero and uniqueness decisions because all contributions
are nonnegative. Repeated local energy values are counted with their
branch indices. A state with count one has a unique predecessor; trace
these predecessors from $C_r(U)=1$. The recurrence gives the stated
time and storage. All combinations are actual by hypothesis, so this
energy count is the actual complete-reply count.
\end{proof}

All-nonzero branches on a cactus, including its bridge coordinates,
satisfy the actualness hypothesis by factor propagation. Branches that
fail local actualness independently of other cycles can first be removed,
as in the reduction of Theorem~\ref{thm:singleton-hardness}. The test also
works with real algebraic roots whose energies are rational and exactly
recognized as such. Large binary weights can make $U$ exponential in
input length. This is a dynamic-programming alternative in a numerical
parameter, whereas meet-in-the-middle has exponential dependence on
the number of branch choices even for binary energies.

\section{A parsimonious reduction and singleton search}\label{app:hardness}

\begin{proof}[Proof of Theorem~\ref{thm:singleton-hardness}]
Start from positive integers $a_1,\ldots,a_r$ and a nonnegative target
$T_{\rm ss}$ in \textsc{Subset Sum} \citep{karp1972}. Put $w_\ell=2a_\ell$,
$p_\ell=(w_\ell+1)/2$, $q_\ell=(w_\ell-1)/2$, and
\[
 c_\ell^-=p_\ell q_\ell-p_\ell-q_\ell,\qquad
 c_\ell^+=p_\ell q_\ell+p_\ell+q_\ell.
\]
In local edge order $(00,01,11,12,22,20)$ define
\[
 V_\ell(t)=(t,1-t,1+t,-t,c_\ell^-+t,c_\ell^+-t).
\]
Its row and column sums are independent of $t$, and its cycle equation is
\begin{equation}
 \prod_{j\text{ even}} V_{\ell,j}(t)
 -\prod_{j\text{ odd}} V_{\ell,j}(t)
 =2t(t-p_\ell)(t-q_\ell).                     \label{eq:subset-cubic}
\end{equation}
At $t=0$, edges $00$ and $12$ vanish while $01$, $11$, $22$, $20$
are nonzero. The endpoints of edge $00$ are both active, so this root
is not actual. Here $c_\ell^-=(w_\ell^2-4w_\ell-1)/4\ne0$ for an even
integer $w_\ell$, and $c_\ell^+>0$.

Both remaining roots give actual edge-nonzero arrays. Indeed, take
$t\in\{p_\ell,q_\ell\}$ and let $z$ be the other root. The factors
\begin{equation}
 a^{\rm loc}(t,z)=\left(1,\frac{1+t}{1-t},\frac{z(1+t)}t\right),\qquad
 b^{\rm loc}(t)=\left(t,1-t,\frac{t(t-1)}{t+1}\right)          \label{eq:subset-factors}
\end{equation}
produce $V_\ell(t)$. Since $p_\ell-q_\ell=1$, the entries on edges
$22$ and $20$ are respectively $z(t-1)$ and $z(t+1)$.
The roots are positive half-integers, neither equal to one, so no factor
or observed entry vanishes.

Glue the first row of every cycle into one vertex; then $m=2r+1$,
$n=3r$. Every actual branch choice has the common first factor one
in \eqref{eq:subset-factors}, hence all $2^r$ choices glue. No choice
using the zero root is actual, by the active-endpoint test above.
These are all gradient candidates, because the alternating kernel
parameters are independent on the flower.

Let $S=\sum_\ell w_\ell$ and $\lambda=1/[10(1+S)]$. Scale both
global factor vectors by $\lambda$, so the observed arrays are
$\lambda^2 V_\ell(t)$. For each root, $t\geq1/2$,
$|(1+t)/(1-t)|\leq5$, $z(1+t)/t\leq3w_\ell$, and every column
factor in absolute value is at most $w_\ell$. Consequently the raw
global squared factor norms are at most
\[
 1+25r+9S^2<100(1+S)^2,\qquad 3S^2<100(1+S)^2.
\]
Thus every choice has strictly subunit factors. These bounds also
give polynomial binary lengths for the factors and public input.

The decisive identity is
\begin{equation}
 \|V_\ell(p_\ell)\|^2-\|V_\ell(q_\ell)\|^2
 =2(p_\ell^2-q_\ell^2)=4a_\ell.                \label{eq:subset-energy}
\end{equation}
Use the fixed legal query $u_i=1/m$, $v_j=1/n$. The gradients are
those of the all-$q_\ell$ array $C_0$. Give the public loss
$f_{C_0}+2\lambda^4T_{\rm ss}$, so its extracted norm is
$\|C_0\|^2+4\lambda^4T_{\rm ss}$. A branch choice agrees with this norm
exactly when the indices choosing $p_\ell$ sum to $T_{\rm ss}$. Distinct choices
give distinct arrays. This proves the polynomial, multiplicity-preserving
reduction and the feasibility lower bound.

For the singleton consequence, suppose a deterministic decoder has
polynomial running time on all promised singleton inputs. Given an
arbitrary \textsc{Subset Sum} instance with $r$ items, independently
choose $c_i\in\{1,\ldots,2r\}$ and set $M=2r^2+1$.
For every $k=0,\ldots,2r^2$, transform the instance with weights
$M a_i+c_i$ and target $M T_{\rm ss}+k$ into the preceding public input.
Run the decoder with its promised polynomial cutoff and accept only
after verifying the recovered subset against the original instance.
Every new solution is an original solution, because
$0\leq\sum_i c_i x_i<M$.

The isolation lemma \citep{mulmuley1987} says that, with probability
at least $1/2$, the original solution family has a unique minimum
$c$-weight subset. The corresponding $k$ then has exactly one solution,
and the promised decoder recovers it. A no-instance is never accepted,
whereas a yes-instance is accepted with probability at least $1/2$.
Thus \textsc{Subset Sum} is in RP, implying $\mathrm{NP}=\mathrm{RP}$.
For completeness, the isolation bound follows by fixing all $c_j$,
$j\ne i$: at most one value of $c_i$ makes the best set containing
$i$ tie the best set excluding it. A nonunique minimum has such a
coordinate, and a union bound gives $r/(2r)=1/2$.
\end{proof}

\section{Bit implementation of the exact predicates}\label{app:bitcomplexity}

We prove Theorem~\ref{thm:bitcomplexity}. All statements here concern
rational public data and an implementation of the finite-list algorithm.
Write $n_0=V+e+1$; empty masks are handled directly over $\mathbb Q$.
For $e>0$, write an edge on cycle $\ell(e)$ as
\[
 C_e=\alpha_e+\beta_e t_{\ell(e)},
\]
An edge outside the cycles is constant, $C_e=\alpha_e$; set $\beta_e=0$
in the coefficient bounds below. The root variables
for different cycles need not generate independent fields. Let $\mathcal N$
be the extracted target norm, and let $Q_{\rm aff}\geq1$ be a positive common
denominator of all $\alpha_e,\beta_e,\mathcal N$. A product of their
denominators suffices.

\paragraph{Input heights and root intervals.}
The proof of Proposition~\ref{prop:height} gives a denominator $Q_{\rm den}^4$
and numerator bit length $O(n_0(b+1))$ for every forest edge. Dividing by
the query weight adds $O(b)$ bits. The same $Q_{\rm den}^4$ clears
$2f+\sum_e u_i^2v_j^2-\inner{u}{g_u}-\inner{v}{g_v}$, whose numerator
also has polynomial bit length $O(n_0(b+1))$.
Thus $\log_2Q_{\rm aff}=O(n_0^2(b+1))$. The integer cycle coefficients have bit
length at most $\tau=O(n_0^2(b+1))$.

For each nonconstant integer cycle polynomial let $a_\ell\geq1$ be the
absolute leading coefficient. Choose a common rational Cauchy bound
$R\geq1$ on all its complex roots with $\log_2R=O(\tau+1)$, for example
$R=2^{\tau+1}$. Square-free isolation and dyadic refinement of its real
roots cost a polynomial in degree, coefficient bits and requested
precision. Standard Sturm--Habicht isolation, including repeated-root
handling, is described by \citet{emiris2008}; see also \citet{basu2006}.
Classical root-separation estimates originate in \citet{mahler1964}
and are developed in \citet{mignotte1992}.
After isolation, each further dyadic bisection uses a Sturm root count
with integers of bit length polynomial in the requested precision.
An interval containing one root therefore needs only a polynomial number
of bit operations to reach any specified $B_{\rm prec}$ bits. The initial isolation
has polynomial bit cost; one does not scan a uniformly fine mesh.

\paragraph{Polynomial predicates after propagation.}
Support zero tests use $C_e$. For nonzero-edge propagation, every factor
is an alternating ratio of products of distinct edges on a simple tree
path, starting one row factor at one. Consistency of factors $N_a/D_a$
and $N_b/D_b$ at an edge is the polynomial zero test
$N_aN_b-C_eD_aD_b=0$; its degree in edge forms is at most $2e+1$.
All path denominators are nonzero because their edges have already
passed the nonzero test. The norm predicate has degree two.

For an actual energy-compatible candidate let $h$ be the number of
nontrivial components of its nonzero-edge graph. Put
\[
 P=\prod_{e:\,C_e\ne0} C_e^2>0,\qquad
 A'_j=PA_j,\quad B'_j=PB_j,\quad R_j=A'_jB'_j.
\]
Here $A_j,B_j$ are the component squared factor norms from
Lemma~\ref{lem:actual}. Each squared path denominator divides $P$.
Consequently $A'_j$ and $B'_j$ are sums of at most $V$ monomials in edge
forms of degree at most $2e+2$, and $R_j$ has degree at most
$L_{\rm pred}:=4e+4$. The optional budget is exactly
\begin{equation}
 S:=\sum_{j=1}^{h}\sqrt{R_j}-P\leq0,\label{eq:cleared-budget}
\end{equation}
using nonnegative square roots. No estimate of a small propagation
denominator is required. These expressions have polynomial size:
one may list path products and at most $V^2$ products of pairs of terms
in each $R_j$, retaining edge forms as linear factors. Expanding into
all monomials in the root variables is unnecessary.

\paragraph{Integral scaling and a nonzero gap.}
Define
\begin{align*}
 Q_{\rm int}&=Q_{\rm aff}^{L_{\rm pred}}\prod_\ell a_\ell^{L_{\rm pred}},\\
 A_{\max}&=\max\{2,|\mathcal N|,\max_e(|\alpha_e|+|\beta_e|R)\},\\
 C_{\max}&=\left\lceil Q_{\rm int}n_0^3A_{\max}^{L_{\rm pred}}\right\rceil.
\end{align*}
These quantities are computable with polynomial bit lengths; specifically
$\log_2Q_{\rm int},\log_2C_{\max}=O(n_0^4(b+1))$ suffices. Every $a_\ell t_\ell$ is
an algebraic integer, regardless of reducibility or multiplicity of the
cycle polynomial. Each constructed predicate polynomial uses products of at most
$L_{\rm pred}$ edge forms, with coefficients $\pm1$, and possibly the scalar
$\mathcal N$. Multiplication by $Q_{\rm int}$ clears their rational denominators
and root leading coefficients. Thus $Q_{\rm int}P$ and $Q_{\rm int}R_j$ are integral.
Also $Q_{\rm int}\sqrt{R_j}$ is integral: its square $Q_{\rm int}^2R_j$ is integral, and
integrality is transitive. Hence $y=Q_{\rm int}S$ is an algebraic integer of degree
at most $D2^h$.

Every conjugate extends to an embedding of a finite field containing the
tuple and the radicals. Under each such embedding every edge form has
modulus at most $A_{\max}$.
The conjugate moduli of $A'_j,B'_j$ are at most $VA_{\max}^{2e+2}$, and those
of $\sqrt{R_j}$ are at most $VA_{\max}^{2e+2}$. Since $h\leq V$, every conjugate
of $y$ has modulus at most $C_{\max}$. If $y\ne0$, its norm from $\mathbb Q(y)$
to $\mathbb Q$ is a nonzero integer. Therefore
\begin{equation}
 |S|\geq\frac1{Q_{\rm int}C_{\max}^{D2^h-1}}\quad(S\ne0).\label{eq:bit-separation}
\end{equation}
For each earlier polynomial predicate $F$, the same $Q_{\rm int}F$ is integral
and its conjugates obey the same uniform bound $C_{\max}$.
Its degree is at most $D$, so \eqref{eq:bit-separation} applies with
$h=0$. This includes false tuples and support-inconsistent arrays;
actualness is only used for the later square-root budget.

\paragraph{Uniform interval precision.}
Let $\operatorname{bits}(z)=\lfloor\log_2z\rfloor+1$ for positive integers
and define
\[
 B_h=\operatorname{bits}(Q_{\rm int})+(D2^h-1)\operatorname{bits}(C_{\max}),
 \qquad \eta_h=2^{-B_h}.
\]
Every nonzero tested expression has modulus at least $\eta_h$. An outward
rational interval of width less than $\eta_h/2$ decides its sign: if it
excludes zero, use its sign; if it contains zero, the expression must
equal zero. This supplies the equality decision without a numerical
tolerance assumption.

For completeness, this width can be reached at polynomial bit cost.
Suppose root intervals have width at most $\epsilon\leq1$ and lie within
$[-R-1,R+1]$. Each edge form has endpoint modulus at most $2A_{\max}$ and
width at most $A_{\max}\epsilon$. A product of at most $L_{\rm pred}$ such intervals has
width at most $L_{\rm pred}(2A_{\max})^{L_{\rm pred}}\epsilon$. Each polynomial above has at most
$n_0^2$ such product terms; the constant norm term has zero width.
Thus one may take $W=\lceil n_0^2L_{\rm pred}(2A_{\max})^{L_{\rm pred}}\rceil$ as a common width
factor for $P,R_j,F$, with $\log_2W$ polynomial in $n_0,b$.
For polynomial tests, root precision $B_0+\log_2W+O(1)$ suffices.
For the budget choose
\[
 \epsilon\leq\frac{\eta_h^2}{64(h+1)^2W}.
\]
Clip a negative lower bound for $R_j$ to zero, and enclose its square
root by outward dyadic bounds with rounding error at most
$\eta_h/(16(h+1))$ at each endpoint. The inequality
$|\sqrt{x}-\sqrt{y}|\leq\sqrt{|x-y|}$ for $x,y\geq0$ shows that the
sum of all square-root widths, plus the width for $P$, is less than
$\eta_h/2$. Thus root precision
$2B_h+O(\log(h+1)+\log W)$ and polynomially many rational operations
suffice. Dyadic square-root bounds use integer square roots on numbers
of polynomial bit length. Products contain at most $L_{\rm pred}$ edge factors,
so intermediate rational numerator and denominator lengths are also
polynomial in this precision and in $n_0,b$.

\paragraph{Enumeration, output and the energy join.}
There are at most $D$ root tuples and polynomially many predicates per
tuple. Energy is tested first. Polynomial predicates use $h=0$, and the
square-root stage is used only on actual energy-compatible arrays, where
$h\leq\kappa$. Isolation, refinement, rational evaluation and printing
affine edge values with indexed isolating-root representations therefore
have the bit bound in \eqref{eq:bitbound}. A nonzero constant cycle
equation terminates earlier. Identically zero cycle equations remain
outside the finite-list claim.

The energy join additionally compares partial sums belonging to two
tuples. Their common field has degree at most $D^2$; independence is not
assumed. Scaling by $Q_{\rm int}^2$ and using conjugate bound $C_{\max}^2$ supplies the
same polynomial-predicate argument with $D^2$ in place of $D$.
There are at most polynomially many comparisons in $D$, and complete
matching blocks contain at most $D$ full tuples. Final budget checks have
the same $\kappa$ bound. Squaring $D$ changes the absolute polynomial
exponent, not the qualitative bound of Theorem~\ref{thm:bitcomplexity}.

\paragraph{Why the bit bound includes support components.}
The optional unit-factor test introduces a sum of square roots even
on a forest. Take
disjoint stars, one row in component $j$, and give its edges rational
values $x_{j1}/T_{\rm budget},\ldots,x_{jd_j}/T_{\rm budget}$, with $T_{\rm budget}>0$ rational.
The gradient margins determine all edges uniquely. Propagation with
row factor one gives $A_j=1$ and
$B_j=\sum_i x_{ji}^2/T_{\rm budget}^2$, so the bounded-class test is precisely
\[
 \sum_j\sqrt{\sum_i x_{ji}^2}\leq T_{\rm budget}.
\]
The rational reply can be constructed directly from these arrays.
Thus sums of rational Euclidean lengths occur even when $r=0$ and
there is only one candidate. This places an additional exact comparison
problem alongside the familiar sum-of-square-roots issues in numerical
complexity \citep{allender2005}; it is not a hardness classification.
Theorem~\ref{thm:bitcomplexity} retains an explicit support-component
parameter for this comparison.

\section{Heights of the separation certificate}\label{app:certificate-heights}
\begin{samepage}
\begin{proposition}[Coefficient heights of the exception certificate]
\label{prop:exception-height}
Use the unscaled reciprocal coordinates $u_i=1/x_i$, $v_j=1/y_j$ of this
subsection. Let $B$ satisfy the target hypothesis of Theorem~\ref{thm:main}
and have rational entries with numerator and denominator bit lengths at
most $\beta\geq1$. Let $e=|E|$, and set
\[
 \mathcal H=(2e+3)(\beta+1)+\lceil\log_2(2e+2)\rceil+2,
 \qquad W=2\lceil\log_2(e+2D)\rceil+1.
\]
For an ordering $\ell_1,\ldots,\ell_s$ of an eligible subset $J$, define
$\mu_0=1$, $h_0=\mathcal H$ and
\begin{equation}
 \mu_i=\mu_{i-1}e_{\ell_i},\qquad
 h_i=e_{\ell_i}h_{i-1}+2\mu_{i-1}\mathcal H
       +\left\lceil\log_2((e_{\ell_i}+2\mu_{i-1})!)\right\rceil.
 \label{eq:exception-height}
\end{equation}
The primitive integer polynomial proportional to $R_J$ has coefficient
bit length at most $h_s$. The primitive version of $P_B$ has bit length
at most
\[
 \max\{1,\ r\mathcal H+(D-1)((r_++1)\mathcal H+r_+W)\}.
\]
These are coefficient-wise target-dependent bounds, separate from the
public-input cycle heights of Proposition~\ref{prop:height}. Substitution
$(x,y)\mapsto(mx,ny)$ for \eqref{eq:grid-query} adds at most
$\lceil L_G\log_2\max\{m,n,1\}\rceil$ to the global bit-length bound.
\end{proposition}
\end{samepage}
The proof in Appendix~\ref{app:exception-height} uses a common target
denominator and the coefficient norm of each Sylvester determinant.

\subsection{Coefficient-height proof}\label{app:exception-height}

\begin{proof}[Proof of Proposition~\ref{prop:exception-height}]
Let $Q_B$ be the product of the denominators of the observed entries of $B$,
so $Q_B\leq2^{e\beta}$. It clears all matching products in every $C_{\ell j}$
and every coefficient of $T_\ell$. For the coefficient $\ell_1$ norm
(including all query variables and any remaining cycle variables), both
$Q_B\widetilde G_\ell$ and $Q_B\sum_{\ell\in J}E_\ell$ have norm less than
$2^{\mathcal H}$. Indeed the two matching expansions have norm at most
$2Q_B(1+2^\beta)^{k_\ell}$, and the energy sum has norm at most
$Q_Bq(1+2^{\beta+1})$; these bounds are strictly below the stated power.

Eliminate using these integral polynomials. At step $i$ the Sylvester
matrix has size $e_{\ell_i}+2\mu_{i-1}$. Each determinant term uses
$2\mu_{i-1}$ coefficients of the cycle polynomial and $e_{\ell_i}$ of
the preceding resultant. Submultiplicativity of the coefficient norm
and the factorial number of determinant terms give
\eqref{eq:exception-height}. This integral resultant is
$Q_B^{M_J+\sum_{\ell\in J}2M_J/e_\ell}R_J$, so primitive normalization
can only reduce its integer coefficient heights.

Put $S_i=e_{\ell_i}+2\mu_{i-1}$ and $S_{\max}=e+2D$.
Since $e_{\ell_i}\geq2$, $\mu_{i-1}\geq1$ and $S_i\leq S_{\max}$,
\[
 \frac{S_i}{\mu_{i-1}e_{\ell_i}}
 =\frac1{\mu_{i-1}}+\frac2{e_{\ell_i}}\leq2.
\]
The elementary bound $S_i!\leq S_i^{S_i}$ therefore gives
\[
 \frac{\lceil\log_2(S_i!)\rceil}{\mu_i}
 \leq 2\log_2 S_{\max}+\frac1{\mu_i}\leq W.
\]
Divide \eqref{eq:exception-height} by $\mu_i=\mu_{i-1}e_{\ell_i}$:
\[
 \frac{h_i}{\mu_i}
 \leq\frac{h_{i-1}}{\mu_{i-1}}
       +\frac{2\mathcal H}{e_{\ell_i}}+W
 \leq\frac{h_{i-1}}{\mu_{i-1}}+\mathcal H+W.
\]
Starting from $h_0/\mu_0=\mathcal H$, induction yields
$h_s\leq M_J((s+1)\mathcal H+sW)$.
Finally the integral leading coefficients have norm below
$2^{\mathcal H}$, and $\sum_{J\neq\varnothing}M_J=D-1$.
Multiplying these bounds proves the assertion for $P_B$ without
expanding it.
\end{proof}

\section{An exact algebraic coverage certificate}\label{app:rur}

The block decoder can represent nonrational candidates by a rational
univariate representation (RUR). We use classical quotient-algebra traces
\citep{rouillier1999}; certified separating forms for nonradical ideals
are also treated in \citet{demin2025rur}. The following elementary
criterion records the completeness checks performed by our implementation.

Let $I=\langle F_1,\ldots,F_s\rangle\subset\mathbb Q[x_1,\ldots,x_k]$
be a proper zero-dimensional norm-augmented reply ideal. Compute a
monomial basis of $\mathcal A=\mathbb Q[x]/I$ and its multiplication
matrices $M_i$; put $\nu=\dim_{\mathbb Q}\mathcal A$. All matrix
identities below are rational. For a rational linear form
$\lambda=\sum_i w_i x_i$, write $M_\lambda=\sum_i w_iM_i$.

\begin{lemma}[RUR coverage check]\label{lem:rur-coverage}
Suppose $P\in\mathbb Q[z]$ is monic and squarefree and
$v_i\in\mathbb Q[z]$ have degrees below $\deg P$. If
\begin{align*}
 F_j(v_1,\ldots,v_k)&\equiv0\pmod P\quad(1\leq j\leq s),
 &\sum_i w_i v_i&\equiv z\pmod P,\\
 P(M_\lambda)^\nu&=0,
 &(M_i-v_i(M_\lambda))^\nu&=0\quad(1\leq i\leq k),
\end{align*}
then $z\mapsto(v_1(z),\ldots,v_k(z))$ bijects the complex roots of
$P$ with $V(I)$ and bijects their real subsets. The ideal need not be
radical.
\end{lemma}
\begin{proof}
At any point $\xi\in V(I)$, evaluation sends nilpotent elements of
$\mathcal A$ to zero. The matrix identities therefore imply
$P(\lambda(\xi))=0$ and $\xi_i=v_i(\lambda(\xi))$. Conversely,
the congruences give a point of $V(I)$ at every root $z$ of $P$, and
$\lambda(v(z))=z$ makes the map injective. Rational coefficients imply
that real roots give real points; a real point has real image under
$\lambda$. These arguments also apply to nonreduced local algebras.
\end{proof}

\paragraph{Deterministic construction.}
Try $\lambda_a=\sum_{i=1}^k a^{i-1}x_i$ at consecutive positive
integers $a$. Two distinct points have equal images for at most $k-1$
values of $a$. Since there are at most $\nu$ distinct points, one of the
first $(k-1)\binom\nu2+1$ forms separates them. For each trial take
$P$ to be the monic squarefree part of the characteristic polynomial of
$M_\lambda$. Write $P(z)=\sum_{h=0}^{d}a_hz^h$, set $M_0$ to the
identity matrix, and form
\[
 Q_j(z)=\sum_{h=1}^{d}a_h\sum_{\ell=0}^{h-1}
       \operatorname{tr}(M_jM_\lambda^\ell)z^{h-1-\ell}
       \quad(0\leq j\leq k).
\]
For a separating trial and a point $\xi$ of local multiplicity
$m_\xi$, the trace on its local algebra is
$\operatorname{tr}_{\xi}(M_jM_\lambda^\ell)
=m_\xi\xi_j\lambda(\xi)^\ell$ (with $\xi_0=1$).
Summing over the local algebras gives
\[
 Q_j(\lambda(\xi))=m_\xi\xi_jP'(\lambda(\xi)),\qquad
 Q_0(\lambda(\xi))=m_\xi P'(\lambda(\xi))\ne0.
\]
Thus $v_i=Q_iQ_0^{-1}\bmod P$ gives the coordinates. For any trial,
$Q_0$ is invertible modulo $P$, with the multiplicities summed over
points sharing a projected value. Test Lemma~\ref{lem:rur-coverage};
a failed nilpotence check requests the next form. A separating form
passes, so this process terminates within the displayed bound.

\paragraph{From roots to complete replies.}
Sturm isolation enumerates every real root of $P$. Substitute the
coordinate polynomials into the affine edge values, verify the original
loss and gradients modulo $P$, and apply Lemma~\ref{lem:actual} in the
corresponding real number field. The optional factor budget uses exact
algebraic zero tests and outward interval refinement. This gives the
complete actual list without a rationality or radical-ideal assumption.
The verifier reconstructs the quotient algebra and checks nilpotence
by matrix powers; the generator uses characteristic polynomials and
traces. A supplied quotient-size or execution budget gives an explicit
\texttt{Unresolved} status when exhausted. For cactus decoding,
Theorem~\ref{thm:bitcomplexity} supplies the deterministic bit bound.

Before constructing the quotient, the implementation rescales flow
variables by the rational gcd of the nonzero weighted forest-base
entries, using scale one when they all vanish, and removes scalar
content from the equations. This invertible public-input preprocessing
preserves reality and multiplicities.

\section*{Code and data availability}
The code, synthetic public inputs, complete computational outcomes and
reproduction instructions are available in software version v1.1.0 at
\url{https://doi.org/10.5281/zenodo.23664952}. Source code:
\url{https://github.com/llljjjwww333/one-loss-rank-one-decoding}.
The software and accompanying synthetic data use the MIT license.
The archive includes budget skips and timeouts, and its full reproduction
command checks the reported outcomes; it does not repeat solver timings.
The manuscript source distribution rebuilds this article and its supplement.

\section*{Declaration of generative AI and AI-assisted technologies
in the manuscript preparation process}
OpenAI Codex assisted with mathematical exploration, proof drafts,
implementation, checking scripts and source-defined plots, as well as
drafting, restructuring and language revision. User-supplied Claude
critiques informed revisions. Figures are reproducible from the reported
data or equations. The author has reviewed and verified the mathematical
arguments, code and manuscript content and takes responsibility for the
work.

\par\medskip
\begingroup
\small
\setlength{\bibsep}{2pt}
\interlinepenalty=10000
\bibliographystyle{plainnat}
\bibliography{references}
\endgroup

\clearpage
\setcounter{section}{0}
\setcounter{table}{0}
\setcounter{figure}{0}
\setcounter{equation}{0}
\renewcommand{\thesection}{S\arabic{section}}
\renewcommand{\thetable}{S\arabic{table}}
\renewcommand{\thefigure}{S\arabic{figure}}
\renewcommand{\theequation}{S\arabic{equation}}
\renewcommand{\theHsection}{supp.\arabic{section}}
\renewcommand{\theHtable}{supp.\arabic{table}}
\renewcommand{\theHfigure}{supp.\arabic{figure}}
\renewcommand{\theHequation}{supp.\arabic{equation}}
\begin{center}\Large\bfseries Supplementary Material\end{center}

This supplementary section specifies the computational protocols,
indexed inputs and complete grid outcomes. The main article, its formal
proof appendices and the bibliography precede this section. Unprefixed
references point to the main article above; supplementary numbers begin with S.

\section{Exact four-cycle grid validation}
The exhaustive cohort comprises six one-cycle grids ($N=1,\ldots,6$),
three two-cycle grids ($N=1,2,3$), and the three-cycle binary grid. Safe points have
nonzero linear cycle equations. At every remaining point, rational
orthogonal reflection constructs a distinct actual candidate with the
same two gradient blocks and total norm. Row-equality witnesses also
preserve both global factor norms.

\begin{table}[ht]\centering\small
\begin{tabular}{rrrrr}\toprule
$r$ & $N$ & Queries & Collisions & Bounded witnesses\\\midrule
1 & 1 & 1 & 1 & 1 \\
1 & 2 & 16 & 12 & 8 \\
1 & 3 & 81 & 45 & 27 \\
1 & 4 & 256 & 112 & 64 \\
1 & 5 & 625 & 225 & 125 \\
1 & 6 & 1296 & 396 & 216 \\
2 & 1 & 1 & 1 & 1 \\
2 & 2 & 128 & 120 & 96 \\
2 & 3 & 2187 & 1755 & 1215 \\
3 & 2 & 1024 & 1008 & 896 \\
\bottomrule\end{tabular}
\caption{Exhaustive rational checks of the four-cycle law. The bounded
column counts the row-equality construction, a lower bound rather than
a complete bounded-fiber classification.}\end{table}

\section{Integer-lattice uniqueness controls}
The rational backend uses only the public mask, query, loss and two
gradients. It solves the flower margins, extracts the rational roots of
each cycle polynomial, rejects locally impossible zero supports, and
computes exact local energies. Denominators are cleared and a common
integer gcd is removed. This rational-only backend reports nonrational roots as unsupported;
the article's general decoder handles algebraic roots.
The three measured combiners all decide the same capped count and return
the sole array when it is unique. Repeated equal-energy branches retain
their multiplicities. The operation counters count full tuples,
compressed partial-sum transitions, or DP recurrence transitions,
respectively; these different elementary operations are mechanism
counts, not wall-time speedup factors.

The enumeration budget is $65536$ full tuples, the meet-in-the-middle
budget is $131072$ tuples per half before compression, and the dense DP
budget is $200000$ stored layer cells. All $32$ public inputs, outcomes,
skips, reconstruction checks and separate timings are reported.

\section{Unified Linux comparison}
The three methods use a common serial protocol: fourteen public
inputs, three repetitions, fixed CPU affinity and a thirty-second
after-ready wall limit. Method order rotates across repetitions.
Public IO, construction, exact root work, native and Python subworker
startup, reconstruction, support checking and serialization are included.
One common parent import is excluded and recorded separately, as are
individual child startup intervals. The host uses Linux/WSL on an AMD
Ryzen 7 5800U, Python 3.12.3, SymPy 1.14.0, msolve 0.6.5 and Singular
4.3.2. Host load is recorded; these are single-host measurements.

\begin{table}[htbp]\centering\small
\begin{tabular}{rlrrr}\toprule
No. & Input & Cycle join & msolve & Edge Singular\\\midrule
1 & 01-r2-k33 & 0.318 & 0.350 & 0.399 \\
2 & 02-r2-k55 & 0.389 & 0.362 & 0.412 \\
3 & 03-r4-k3333 & 0.320 & 0.357 & 0.744 \\
4 & 04-r4-k5555 & 0.677 & 0.444 & 0/3; TO \\
5 & 05-r8-k33333333 & 0.578 & 8.545 & 0/3; TO \\
6 & 06-r8-k55555555 & 2.533 & 0/3; TO & 0/3; TO \\
7 & 07-r12-k333333333333 & 0.917 & 0/3; TO & 0/3; TO \\
8 & 08-r12-k555555555555 & 10.208 & 0/3; TO & 0/3; TO \\
9 & 09-chain-r3-k234 & 0.259 & 0.358 & 0.415 \\
10 & 10-bridge-r3-k234 & 0.274 & 0.354 & 0.426 \\
11 & 11-disjoint-r3-k234 & 0.327 & 0.353 & 0.415 \\
12 & 12-zero\_path-r3-k234 & 0.255 & 0.357 & 0.426 \\
13 & two\_C6\_N1\_0 & 0.214 & 0.356 & 0.447 \\
14 & C8\_N4\_83 & 0.245 & 0.343 & 0.444 \\
\bottomrule\end{tabular}
\caption{Seconds under one protocol. A reported time is the median of
three completed runs; otherwise the cell states completed runs out of
three and TO marks the thirty-second censoring limit. Timeouts are not
substituted as measured runtimes.}\label{tab:strengthened-times}
\end{table}

The raw files include every run, startup interval, load record, generated
native system and complete solver outcome. The cycle system and native
observed-edge system encode the same reply; the latter constructs no
shared symbolic incidence reduction. Every completed comparison list
is checked with rational arithmetic against the original loss and
gradients and against the complete reference list.

The separate phase protocol uses the same machine and affinity after
the comparison finishes. It measures decode calls on all ninety-six
prelisted inputs, with three fresh fork children per input. Imports,
public JSON parsing and final file writing lie outside these phase
times. Those times are not pooled with the end-to-end comparison.

\begin{table}[htbp]\centering\small
\begin{tabular}{rrrrrrrr}\toprule
$r$ & Cert./16 & Total & Reduction & Roots & Coord. & Prop. & Check/other\\\midrule
1 & 16 & 0.195647 & 0.192700 & 0.001686 & 0.000017 & 0.000067 & 0.001177 \\
2 & 16 & 0.198293 & 0.193494 & 0.002678 & 0.000030 & 0.000110 & 0.001982 \\
4 & 16 & 0.213099 & 0.204589 & 0.004655 & 0.000048 & 0.000265 & 0.003542 \\
8 & 13 & 0.248294 & 0.230956 & 0.009329 & 0.000114 & 0.000834 & 0.007061 \\
12 & 15 & 0.269140 & 0.244277 & 0.012960 & 0.000185 & 0.001630 & 0.010088 \\
24 & 8 & 0.347339 & 0.296533 & 0.025150 & 0.000581 & 0.006289 & 0.018787 \\
\bottomrule\end{tabular}
\caption{Coordinate-decoder phase times in seconds. For each input, take
the median-total run and its corresponding phases, then average over
the sixteen inputs at each size. Check/other includes final rational
reply/support checking, local energy evaluation and result assembly.
Reduction includes rational public-data canonicalization, incidence
solving and cycle-equation construction. All outcomes agree across
the three repetitions.}
\label{tab:coordinate-phases}\end{table}

\begin{table}[htbp]\centering\small
\begin{tabular}{rrrrrr}\toprule
Input & Real tuples & Left list & Right list & Comparisons & Matches\\\midrule
1 & 9 & 3 & 3 & 12 & 1 \\
2 & 9 & 3 & 3 & 9 & 1 \\
3 & 27 & 9 & 3 & 30 & 1 \\
4 & 225 & 15 & 15 & 101 & 1 \\
5 & 6561 & 81 & 81 & 902 & 1 \\
6 & 140625 & 375 & 375 & 5488 & 1 \\
7 & 59049 & 243 & 243 & 3362 & 1 \\
8 & 6834375 & 3375 & 2025 & 44324 & 1 \\
9 & 3 & 3 & 1 & 4 & 1 \\
10 & 9 & 3 & 3 & 12 & 1 \\
11 & 9 & 3 & 3 & 10 & 1 \\
12 & 3 & 3 & 1 & 6 & 1 \\
13 & 9 & 3 & 3 & 14 & 2 \\
14 & 3 & 3 & 1 & 7 & 2 \\
\bottomrule\end{tabular}
\caption{Complete cycle-join operation counts on the same fourteen inputs.
Real tuples count the Cartesian product before the energy test; the two
partial lists and comparisons are actually generated by the exact join.
Matches count full root tuples satisfying energy equality before support or budget tests. These operation counts are separate from measured seconds.}
\label{tab:join-operations}\end{table}

\FloatBarrier

\section{Input definitions and certificate representation}
The unified comparison uses twelve main inputs and two nonsingleton
controls. The public data index identifies each array and its file hash.
Input \texttt{validation-06-}\allowbreak\texttt{flower-r8-k55555555}
has $140625$ real root tuples and two half lists of $375$.
Cycle lengths and cycle count alone do not specify an input: distinct
arrays with the same $(r,k)$ can have $30375$ or $50625$ root tuples.
The comparison table uses the fourteen indexed arrays and one timing
protocol throughout.

The certificate experiments distinguish symbolic representation from
full expansion. The exception polynomial has explicit degree and
coefficient-height bounds. Determinant and resultant representations specify it without
full symbolic expansion. Four two-cycle expansion jobs reached their
$90$-second limits. Single-cycle height bounds of $569$ and $1570$ bits
correspond to observed heights of $15$ and $37$ bits; these are arithmetic
upper bounds. Algorithm~1 works from public inputs without expanding
the target-dependent certificate.

\section{Larger flowers and small reciprocal grids}\label{sec:large-grid}
Targets alternate the raw arrays $P,M$ of \eqref{eq:PM}, scaled by $1/(256r^2)$ on each cycle. The shared raw row factor is one; dividing all global factors by $16r$ gives strict unit-factor bounds. Queries use \eqref{eq:grid-query}, and candidate fibers contain all actual real arrays without factor bounds.
Public cycle cubics factor into linear and quadratic factors. A nonzero public margin on the shared row ensures that actual local branches glue. Rational-square ratio tests group quadratic energies by positive square class. Distinct classes, together with one, are linearly independent over $\mathbb Q$: character projections in their multiquadratic extension isolate each coefficient. Energy equality is thus rational-vector equality; half-list matching preserves multiplicity.
For every $N>1$, the $64$ query vectors are independent uniform
draws \emph{with replacement} from the full reciprocal grid, whose
coordinates are sampled independently. Repeated query vectors count as
separate draws. In particular, at $r=1,N=2$ the grid has $64$ points,
but the $64$ recorded draws visit $39$ distinct points; this is not an
exhaustive grid. The unique point at $N=1$ is evaluated once.
All $1605$ replies resolve. Four controls agree with the general decoder; rational exhaustions check $81$ and $6561$ tuples at $r=4,8$, $N=1$, and $256$ symbolic controls check radical signatures. All $65$ complete-reply collisions are included.
\begingroup\small
\begin{longtable}{rrrrrrr}
\caption{Larger-flower exact counts. $A_g,A_{g,f}$ count nonsingleton fibers. At $N>1$, $n=64$ counts draws with replacement and the intervals are pointwise $95\%$ Wilson intervals; a dash marks the exhaustive one-point grid at $N=1$. The last column is the largest complete-list size observed in that cell.}\label{tab:large-grid-counts}\\
\toprule $r$ & $N$ & $n$ & $A_g$ & $A_{g,f}$ & Interval & Max. list\\\midrule\endfirsthead
\toprule $r$ & $N$ & $n$ & $A_g$ & $A_{g,f}$ & Interval & Max. list\\\midrule\endhead
\bottomrule\endfoot
1 & 1 & 1 & 1 & 0 & -- & 1 \\
1 & 2 & 64 & 55 & 5 & $[0.034,0.170]$ & 2 \\
1 & 4 & 64 & 45 & 0 & $[0.000,0.057]$ & 1 \\
1 & 8 & 64 & 45 & 0 & $[0.000,0.057]$ & 1 \\
1 & 16 & 64 & 38 & 0 & $[0.000,0.057]$ & 1 \\
1 & 32 & 64 & 45 & 0 & $[0.000,0.057]$ & 1 \\
2 & 1 & 1 & 1 & 1 & -- & 2 \\
2 & 2 & 64 & 58 & 4 & $[0.025,0.150]$ & 2 \\
2 & 4 & 64 & 53 & 0 & $[0.000,0.057]$ & 1 \\
2 & 8 & 64 & 60 & 0 & $[0.000,0.057]$ & 1 \\
2 & 16 & 64 & 58 & 0 & $[0.000,0.057]$ & 1 \\
2 & 32 & 64 & 59 & 0 & $[0.000,0.057]$ & 1 \\
4 & 1 & 1 & 1 & 1 & -- & 6 \\
4 & 2 & 64 & 64 & 5 & $[0.034,0.170]$ & 2 \\
4 & 4 & 64 & 64 & 0 & $[0.000,0.057]$ & 1 \\
4 & 8 & 64 & 63 & 0 & $[0.000,0.057]$ & 1 \\
4 & 16 & 64 & 64 & 0 & $[0.000,0.057]$ & 1 \\
4 & 32 & 64 & 64 & 0 & $[0.000,0.057]$ & 1 \\
8 & 1 & 1 & 1 & 1 & -- & 70 \\
8 & 2 & 64 & 64 & 21 & $[0.226,0.450]$ & 4 \\
8 & 4 & 64 & 64 & 0 & $[0.000,0.057]$ & 1 \\
8 & 8 & 64 & 64 & 1 & $[0.003,0.083]$ & 2 \\
8 & 16 & 64 & 64 & 0 & $[0.000,0.057]$ & 1 \\
8 & 32 & 64 & 64 & 0 & $[0.000,0.057]$ & 1 \\
12 & 1 & 1 & 1 & 1 & -- & 924 \\
12 & 2 & 64 & 64 & 25 & $[0.281,0.513]$ & 55 \\
12 & 4 & 64 & 64 & 0 & $[0.000,0.057]$ & 1 \\
12 & 8 & 64 & 64 & 0 & $[0.000,0.057]$ & 1 \\
12 & 16 & 64 & 64 & 0 & $[0.000,0.057]$ & 1 \\
12 & 32 & 64 & 64 & 0 & $[0.000,0.057]$ & 1 \\
\end{longtable}\endgroup
For $64$ draws and zero collisions, the pointwise Wilson upper endpoint is about $5.7\%$. The sufficient bound becomes nonvacuous only at $N\geq11,69,1289,209937,25509145$ for the five cycle counts. Grid collision probabilities need not be monotone; the $r=8,N=8$ event is included. These observations do not give a smaller uniform bound.

\section{Finite-grid observation-channel comparison}\label{sec:query-probability-details}
The three targets and all queries are specified before the experiment. On $C_6$, use raw factors $(1,1,1)$ and $(1,1,-2)$, each divided by eight. On $C_8$, use $(1,2,3,4)$ and $(2,-1,3,-2)$, each divided by sixteen. The local edge order is $(j,j), (j+1\bmod k,j)$ for $j=0,\ldots,k-1$. The two-cycle target is exactly \eqref{eq:PM}.
For each target and $N>1$, we take $128$ independent uniform
query vectors \emph{with replacement}; each coordinate is sampled
independently from $\{1,\ldots,N\}$. The count $n$ includes repetitions.
For the single six-cycle at $N=2$, the $128$ draws visit $57$ of the
$64$ possible points. At $N=1$ the unique point is evaluated once.
Each public reply is decoded without the target. A forest particular solution gives cycle polynomials of degree at most three. Rational factorization supplies linear and quadratic factors; exact quadratic-field arithmetic tests total energy. The support criterion in Lemma~\ref{lem:actual} filters nonactual arrays. Targets enter only the final retention check. The general decoder cross-checks the first query for each target at $N=1,2$ and every observed complete-reply collision. The raw records retain every candidate and decision.
\begin{longtable}{llrrr}
\caption{Exact finite-grid counts. $n$ counts draws with replacement at $N>1$ and the unique grid point at $N=1$; $A_g$ and $A_{g,f}$ count nonsingleton gradient and complete fibers. No query is unresolved.}\label{tab:query-probability-counts}\\
\toprule Target & $N$ & $n$ & $A_g$ & $A_{g,f}$\\\midrule\endfirsthead
\toprule Target & $N$ & $n$ & $A_g$ & $A_{g,f}$\\\midrule\endhead
\bottomrule\endfoot
One $6$-cycle & 1 & 1 & 1 & 0 \\
One $6$-cycle & 2 & 128 & 115 & 0 \\
One $6$-cycle & 4 & 128 & 126 & 0 \\
One $6$-cycle & 8 & 128 & 123 & 0 \\
One $6$-cycle & 16 & 128 & 124 & 0 \\
One $6$-cycle & 32 & 128 & 127 & 0 \\
One $6$-cycle & 64 & 128 & 125 & 0 \\
One $6$-cycle & 128 & 128 & 125 & 0 \\
One $6$-cycle & 256 & 128 & 120 & 0 \\
One $8$-cycle & 1 & 1 & 1 & 0 \\
One $8$-cycle & 2 & 128 & 128 & 0 \\
One $8$-cycle & 4 & 128 & 119 & 1 \\
One $8$-cycle & 8 & 128 & 118 & 0 \\
One $8$-cycle & 16 & 128 & 117 & 0 \\
One $8$-cycle & 32 & 128 & 120 & 0 \\
One $8$-cycle & 64 & 128 & 119 & 0 \\
One $8$-cycle & 128 & 128 & 116 & 0 \\
One $8$-cycle & 256 & 128 & 115 & 0 \\
Two $6$-cycles & 1 & 1 & 1 & 1 \\
Two $6$-cycles & 2 & 128 & 120 & 7 \\
Two $6$-cycles & 4 & 128 & 114 & 0 \\
Two $6$-cycles & 8 & 128 & 116 & 0 \\
Two $6$-cycles & 16 & 128 & 121 & 0 \\
Two $6$-cycles & 32 & 128 & 110 & 0 \\
Two $6$-cycles & 64 & 128 & 111 & 0 \\
Two $6$-cycles & 128 & 128 & 116 & 0 \\
Two $6$-cycles & 256 & 128 & 116 & 0 \\
\end{longtable}
These fixed-target experiments compare observation channels under the specified discrete query laws. Grid distributions need not have monotone collision probabilities. The continuous almost-sure assertion follows from the analytic theorem; the sampled grid frequencies describe the finite-grid behavior.

\end{document}